\documentclass[12pt]{article}

\usepackage{epsfig}
\usepackage{amssymb}
\usepackage{axodraw}

\newcommand{\beq}{\begin{equation}}
\newcommand{\bea}{\begin{eqnarray}}
\newcommand{\eeq}{\end{equation}}
\newcommand{\eea}{\end{eqnarray}}

\newcommand{\bp}{{\mathbf p}}

\begin{document}

\title{\bf Nonthermal fixed points and the functional renormalization group}

\author{
J{\"u}rgen Berges, Gabriele Hoffmeister\\[0.5cm]
Institute for Nuclear Physics\\
Darmstadt University of Technology\\
Schlossgartenstr. 9, 64289 Darmstadt, Germany}

\date{}
\begin{titlepage}
\maketitle
\def\thepage{}          

\begin{abstract}
\end{abstract}
Nonthermal fixed points represent basic properties of quantum field theories, in addition to vacuum or thermal equilibrium fixed points. The functional renormalization group on a closed real-time path provides a common framework for their description. For the example of an $O(N)$ symmetric scalar theory it reveals a hierarchy of fixed point solutions, with increasing complexity from vacuum and thermal equilibrium to nonequilibrium.  
\end{titlepage}

\renewcommand{\thepage}{\arabic{page}}

\section{Introduction}

Thermal equilibrium properties of many-body or field theories are known to be efficiently classified in terms of renormalization group fixed points. A particularly powerful concept is the notion of infrared fixed points, which are characterized by universality. These correspond to critical phenomena in thermal equilibrium, where a characteristic large correlation length leads to independence of long-distance properties from details of the underlying microscopic theory. In contrast, a classification of properties of theories far from thermal equilibrium in terms of renormalization group fixed points is much less developed. The notion of universality or critical phenomena far from equilibrium is to a large extent unexplored, in particular, in relativistic quantum field theories. Here the strong interest is mainly driven by theoretical and experimental advances in our understanding of early universe cosmology as well as relativistic collision experiments of heavy nuclei in the laboratory. 

It has been recently demonstrated that nonthermal fixed points exist in relativistic self-interacting scalar field theories, and that the associated far-from-equilibrium long-distance properties are characterized by universality~\cite{Berges:2008wm}. Compared to a thermal distribution at low momenta $\sim p^{-1}$, a nonthermal scaling solution was shown to exhibit strongly enhanced fluctuations $\sim p^{-4}$ in the infrared for three spatial dimensions. From a large class of initial conditions leading to nonequilibrium instabilities, such as spinodal decomposition or parametric resonance, these nonthermal fixed points are approached without fine-tuning of parameters. As a consequence, they may have an important impact on properties of early universe dynamics after inflation, or on the physics of thermalization in heavy-ion collisions. 

In this work we employ the functional renormalization group for the average action~\cite{Wetterich:1992yh} to determine properties of nonthermal fixed points for $O(N)$ symmetric scalar quantum field theories. This provides a complementary approach to the two-particle irreducible (2PI) effective action techniques that were used in Ref.~\cite{Berges:2008wm}. The functional renormalization group has a long record of successful applications to the nonperturbative physics of critical phenomena in thermal equilibrium as reviewed in Refs.~\cite{Reviews,Pawlowski:2005xe}. While in equilibrium typically a Euclidean formulation is adequate, nonequilibrium properties require real-time descriptions. For quantum systems a generating functional for nonequilibrium correlation functions with given density matrix at initial time can be written down using the Schwinger-Keldysh closed time path contour following standard procedures~\cite{Berges:2004yj}. In principle, this can be used to construct a nonequilibrium functional renormalization group along similar lines as for Euclidean field theories in thermal equilibrium. However, important differences include the absence of time-translation invariance for general out-of-equilibrium situations~\cite{Gasenzer:2007za}. The nonequilibrium renormalization group takes on a particularly simple form at a nonthermal fixed point, since it becomes independent of the initial density matrix and we concentrate on this in the following. 

Far-from-equilibrium critical phenomena can be observed in many systems, ranging from self-organized criticality in the well-known Bak-Tang-Wiesenfeld sandpile model~\cite{Sandpile} to phase transitions between nonequilibrium stationary states for species dynamics in biologically motivated 
descriptions \cite{Bio}. Renormalization group techniques have been employed to nonequilibrium critical dynamics in a variety of formulations, with a large body of work concerning applications to classical reaction-diffusion problems~\cite{reaction-diffusion,nonpert-reac-diff}, stationary-state transport through quantum systems coupled to reservoirs~\cite{Kehrein,Mitra,Gezzi,Jakobs}, or cosmology~\cite{Calzetta:1999zr}. 

Our paper is organized as follows. Sec.~\ref{sec:introfp} gives an introduction to nonthermal fixed points. Sec.~\ref{sec:ERG} discusses the exact renormalization group on a closed time path, where \ref{sec:genfunc} describes the connection to the 2PI effective action for later comparison. Though the derivation is in principle a straightforward generalization of well-known vacuum or thermal equilibrium techniques as pointed out in Refs.~\cite{nonpert-reac-diff,Gezzi,Jakobs}, there are some important differences. This concerns the exact renormalization group equation for the scale-dependent generating functional and its relation to standard diagrammatic expressions. From the functional form of the renormalization group equation given in Sec.~\ref{sec:exactflow} the flow of all correlation functions can be obtained by further functional differentiation. In contrast, the standard nonequilibrium diagrammatic rules derived in Sec.~\ref{sec:dia} exploit the vanishing of anomalous propagators and vertices at extrema of the generating functional. Another important difference is the apparently larger freedom of choosing cutoff functions on a closed time path. Sec.~\ref{sec:exactflow} discusses constraints which arise from constructing a proper renormalization group flow that interpolates between short and long distance scales. The analysis of nonthermal fixed points is given in Sec.~\ref{sec:fixedpoints}. Sec.~\ref{sec:scaling} describes the scaling form of correlation functions, which are evaluated from the flow equations using a resummed large-$N$ expansion to next-to-leading order (NLO) in Sec.~\ref{sec:2PIlargeN}. The fixed point results are discussed in Sec.~\ref{sec:fixedpointscalc}. Conclusions are given in Sec.~\ref{sec:conclusions}. Two Appendices discuss details about the functional form of propagators and a specific representation for cutoff functions.

\subsection{Thermal versus nonthermal fixed points}
\label{sec:introfp}

Important questions about the evolution of the post-inflationary early universe or heavy-ion collision experiments concern thermalization processes. Starting from a given state or density matrix at some initial time, one considers the nonequilibrium time evolution towards thermal equilibrium. The time to approach the latter can crucially depend on the presence of infrared fixed points at which time scales diverge. This is a well known phenomenon near thermal second-order phase transitions where the dynamics is characterized by critical slowing down. Nonthermal fixed points can have the dramatic consequence that a diverging time scale exists far from equilibrium which prevents or substantially delays thermalization, depending on how closely the fixed point is approached during the nonequilibrium time evolution. 

Fixed points correspond to time and space translation invariant scaling solutions for correlation functions. Here we consider a relativistic scalar field theory with Heisenberg field operator $\Phi(x)$.
For instance, the two-point correlation function $G(x,y)$ can be identically written as
\begin{eqnarray}
G(x,y) &=& \langle \Phi(x) \Phi(y) \rangle \Theta(x^0-y^0) + \langle \Phi(y) \Phi(x) \rangle \Theta(y^0-x^0) \nonumber\\
&=& F(x,y) -\frac{i}{2} \rho(x,y) \, {\rm sign}(x^0-y^0) \, ,
\end{eqnarray}
where $\langle \ldots \rangle$ involves the trace over a given density matrix and $x=(x^0,\bf{x})$ with time $x^0$ and vector ${\bf x}$ in $d$ spatial dimensions~\cite{Berges:2004yj}. The statistical two-point function $F(x,y)$ determines the anti-commutator, while the spectral function $\rho(x,y)$ describes the commutator of two fields according to    
\begin{equation}
F(x,y) = \frac{1}{2}\langle \{ \Phi(x),\Phi(y) \} \rangle \quad , \quad
\rho(x,y) = i \langle [\Phi(x),\Phi(y)] \rangle \, . 
\label{eq:Frhodef}
\end{equation}          
Loosely speaking, the spectral function $\rho(x,y)$ determines which states are available, while the statistical propagator $F(x,y)$ contains the information about how often a state is occupied. 

A tremendous simplification of thermal equilibrium is that the spectral and statistical two-point functions are related by the fluctuation-dissipation relation. In Fourier space with momentum $p$ the latter reads
\begin{equation}
F(p)|_{\rm thermal\,\, equilibrium} = - i\left( n_{T}(p^0) + \frac{1}{2} \right) \rho(p)|_{\rm thermal\,\, equilibrium} \, ,
\label{eq:flucdiss}
\end{equation}    
where $n_{T}(p^0) = (\exp(p^0/T) - 1)^{-1}$ denotes the Bose-Einstein distribution for a temperature $T$. In the infrared this distribution
is given by
\begin{equation}
n_{T}(p^0)\,\, \stackrel{p^0\ll T}{\sim}\,\, \frac{T}{p^0} \, .
\end{equation}

In this work, we consider scaling behavior at an infrared fixed point that can be expressed in terms of the standard dynamical scaling exponent $z$, the anomalous dimension $\eta$ and an "occupation number exponent" $\kappa$ according to\footnote{In Ref.~\cite{Berges:2008wm} this exponent is called $\alpha$ following literature on ultraviolet (UV) fixed points related to turbulence (see also Sec.~\ref{sec:fixedpointscalc}). Since here we concentrate on infrared fixed points associated to critical phenomena we use $\kappa$, which cannot be confused with the specific heat exponent.} 
\begin{equation}
F(p^0,\bp) = s^{2+\kappa} F(s^z p^0, s\bp) \quad , \quad
\rho(p^0,\bp) = s^{2-\eta} \rho(s^z p^0, s\bp) \, .
\label{eq:scaling}
\end{equation}
Inserting (\ref{eq:scaling}) into (\ref{eq:flucdiss}) leads to the following relations for 
\begin{eqnarray}
\mbox{vacuum } (T=0): && \kappa \,=\, - \eta \, ,
\label{eq:vacuumconst}
\\
\mbox{thermal equilibrium } (T\neq 0): && \kappa \,=\, - \eta + z \, .
\label{eq:constraint}
\end{eqnarray}
In contrast, out of equilibrium $F(p)$ and $\rho(p)$ are not constrained by a fluctuation-dissipation relation. This additional freedom allows new scaling solutions, which do not exist in thermal equilibrium because of the presence of (\ref{eq:flucdiss}). We show in Sec.~\ref{sec:fixedpoints} that one can find at a 
\begin{eqnarray}
\mbox{nonthermal fixed point} : && \kappa \,=\, - \eta + z + d\, ,
\label{eq:nonthermalconstraint}
\end{eqnarray}
using nonperturbative large-$N$ expansion techniques to NLO. In contrast to the equilibrium relations (\ref{eq:vacuumconst}) and (\ref{eq:constraint}), out of equilibrium the 
dimensionality of space $d$ enters in (\ref{eq:nonthermalconstraint}). This leads to comparably large values of $\kappa$, which corresponds to strongly enhanced infrared fluctuations as observed in Ref.~\cite{Berges:2008wm}.
As for the thermal equilibrium case, we show that nonthermal scaling solutions correspond to fixed points of the functional renormalization group. In order to observe both thermal and nonthermal fixed points the functional renormalization group has to be considered on a closed time path, which implements the possible absence of a fluctuation-dissipation relation.

\section{Functional RG on a closed time path}
\label{sec:ERG}

\subsection{Generating functional}
\label{sec:genfunc}

We consider a scalar $N$-component quantum field theory with $O(N)$ symmetry, whose classical action reads 
\begin{equation}
  S[\varphi]
  = \frac{1}{2}\int_{x y,{\cal C}} \varphi_a(x) iD_{ab}^{-1}(x,y)\varphi_b(y)
  -\frac{\lambda}{4!N} \int_{x,{\cal C}} \varphi_a(x)\varphi_a(x)\varphi_b(x)\varphi_b(x) . 
  \label{eq:Sq}
\end{equation}
Here the free classical inverse propagator is given by
\begin{equation}
i D^{-1}_{ab}(x,y) = - \left( \square_x + m^2 \right) \delta_{ab}
\delta^{(d+1)}(x-y) \, , 
\label{eq:classprop}
\end{equation}
with $\square_x = \partial^2_{x^0} - \partial^2_{\bf x}$. 
Summation over repeated indices $a=1,\ldots,N$ is implied and
we use the shorthand notation $\int_{x,{\cal C}} \equiv
\int_{\cal C} {\mathrm d} x^0 \int {\mathrm d}^d x$.
For nonequilibrium initial-value problems, the closed time path $\cal C$ leads from
the initial time $t_0$ along the real time axis and back to $t_0$~\cite{CTP}. For the discussion of stationary, i.e.\ time translation invariant nonequilibrium solutions $t_0 \to -\infty$, which we will consider in the following.

We follow closely Refs.~\cite{Berges:2007ym,Chou:1984es} 
and employ a standard notation, where the superscripts `$+$' and `$-$'
indicate that the fields are taken on the
forward (${\cal C}^+$) and backward (${\cal C}^-$) branch of the
closed time path, respectively. Then the action
(\ref{eq:Sq}) can be written as
\begin{eqnarray}
  S[\varphi^+,\varphi^-]\!\!
  &=&\!\!
  \frac{1}{2}\int_{xy}\!\! \left( \varphi^+_a(x),\varphi^-_a(x) \right)
  \left(\! \begin{array}{cc}
    iD^{-1}_{ab}(x,y)   & 0\\
    0           &  -iD^{-1}_{ab}(x,y)
  \end{array}\!\right)
  \left(\!\begin{array}{c} \varphi^+_b(y) \\ \varphi^-_b(y)
  \end{array}\!\right)
  \nonumber\\
  &-& \frac{\lambda}{4!N} \int_{x} \left(
   \varphi_a^+(x)\varphi_a^+(x)\varphi_b^+(x)\varphi_b^+(x)
  -\varphi_a^-(x)\varphi_a^-(x)\varphi_b^-(x)\varphi_b^-(x)\right) ,
  \nonumber\\
\label{eq:Spm}
\end{eqnarray}
where the minus sign in front of the `$-$' terms accounts
for the reversed time integration with $\int_{x} \equiv
\int_{-\infty}^{\infty} {\mathrm d} x^0 \int {\mathrm d}^d x \equiv \int {\mathrm d}^{d+1} x$.
The corresponding generating functional for correlation functions then reads:
\begin{eqnarray}
\label{eq:NEgenFuncZphipm}
  Z[J,R] \!\!&=&\!\! 
     \int {\cal D}\varphi^+{\cal D}\varphi^-
     \exp\Bigg\{i\Bigg[S[\varphi^+,\varphi^-]
   + \int_{x}\,\left( \varphi^+_a(x), \varphi^-_a(x)\right)\left(\!\begin{array}{c} J^+_a(x) \\ - J^-_a(x)
  \end{array}\!\right)
  \nonumber\\
  &+&\!\! \frac{1}{2}\int_{xy}\!\! \left( \varphi^+_a(x),\varphi^-_a(x)
  \right)\left(\!\begin{array}{cc} R^{++}_{ab}(x,y) & -R^{+-}_{ab}(x,y)\\
  -R^{-+}_{ab}(x,y) & R^{--}_{ab}(x,y) \end{array}\!\right)
  \left(\!\begin{array}{c} \varphi^+_b(y) \\ \varphi^-_b(y)
  \end{array}\!\right)
\Bigg]\Bigg\}.
\nonumber\\
\end{eqnarray}
Here we introduced apart from the source terms linear in the fields, $J^+$ and $J^-$, also sources bilinear in the fields, $R^{++}$ etc. Again the superscripts `$+$' and `$-$' indicate that the respective time arguments of the sources are taken on the forward or backward branch of the
closed time path. As a consequence of this notation all space-time integrations run from $-\infty$ to $\infty$, thus calculations are done in $d+1$-dimensional Minkowski space-time. 

The connection to statistical and spectral components as employed in Sec.~\ref{sec:introfp} is conveniently obtained with a linear transformation $A$ of the fields 
\begin{eqnarray}
  \left(\begin{array}{c}
  \varphi \\
  {\tilde\varphi}
  \end{array}\right)
  &=\, A\,\left(\begin{array}{c}
  \varphi^+ \\
  \varphi^-
  \end{array}\right),
\label{eq:RTransf}
\end{eqnarray}
where
\begin{eqnarray}
  &&A
  = \left(\begin{array}{rr} \frac{1}{2} & \frac{1}{2} \\
                             1           & -1
                             \end{array}\right), \qquad
  A^{-1}
  = \left(\begin{array}{rr} 1 & \frac{1}{2} \\
                             1 & -\frac{1}{2}
                             \end{array}\right).
\label{eq:RMatrix}
\end{eqnarray}
We note that $\varphi$ agrees with the defining field in (\ref{eq:Sq}) only for $\varphi^+ =
\varphi^-$. Since this will be the case for expectation values
in the absence of sources and since there is no danger of confusion in the
following we keep this notation.
Correspondingly, we write for the source terms
\begin{equation}
  \left(\begin{array}{c} J \\ \tilde{J} \end{array}\right)
  = A\,\left(\begin{array}{c} J^+ \\ J^-
  \end{array}\right) ,\,\,
 \left(\begin{array}{rr}
  R^F & R^{\mathrm R} \\
  R^{\mathrm A} & R^{\tilde F}
  \end{array}\right)
  = A
  \left(\begin{array}{rr}
  R^{++}  & R^{+-} \\
  R^{-+}  & R^{--}
  \end{array}\right)A^{T} .
\label{eq:KMatrix}
\end{equation}
Inserting these definitions\footnote{Note that
(\ref{eq:KMatrix}) can be equivalently written as
\begin{eqnarray}
 \left(\begin{array}{c}
  \tilde{J} \\ J
  \end{array}
  \right)
  = \left(A^{-1}\right)^T
  \left(\begin{array}{c}
  J^+ \\ - J^-
  \end{array}\right) \quad ,\quad
 \left(\begin{array}{rr}
  R^{\tilde F}\!     &\! R^{\mathrm A} \\
  R^{\mathrm R}\!      &\! R^F
  \end{array}\right)
  = \left(A^{-1}\right)^T
  \left(\begin{array}{rr}
  R^{++}\!  &\! -R^{+-} \\
  -R^{-+}\! &\! R^{--}
  \end{array}\right)A^{-1}.\nonumber
\end{eqnarray}}
into the functional integral
(\ref{eq:NEgenFuncZphipm}) yields
\begin{eqnarray}
  Z[J,\tilde{J},R^{\mathrm{R},\mathrm{A},F,{\tilde
  F}}] \!\!&=&\!\! \int {\cal D}\varphi{\cal D} {\tilde\varphi}
     \exp\Bigg\{i\Bigg[S[\varphi,{\tilde\varphi}]
     + \int_x \left( \varphi_a(x), {\tilde\varphi}_a(x)\right)
               \left(\!\begin{array}{c}{\tilde J}_a(x) \\ J_a(x)\end{array}\!\right)
  \nonumber\\
  &+&\!\! \frac{1}{2}\int_{xy}\!\!
  \left( \varphi_a(x),{\tilde\varphi}_a(x) \right)
  \left(\!\begin{array}{cc}
  R^{\tilde F}_{ab}(x,y)         &\!\! R^\mathrm{A}_{ab}(x,y)\\[1pt]
  R^\mathrm{R}_{ab}(x,y)         &\!\! R^{F}_{ab}(x,y)
  \end{array}\!\right)
  \left(\!\begin{array}{c} \varphi_b(y) \\ {\tilde\varphi}_b(y)\end{array}\!\right)
  \!\Bigg]\!\Bigg\},
  \nonumber\\
\label{eq:ZQuantPhiXi}
\end{eqnarray}
where $S[\varphi,{\tilde\varphi}] = S_0[\varphi,{\tilde\varphi}] +
S_{\rm int}[\varphi,{\tilde\varphi}]$ consists of the action for
the free field theory
\begin{equation}
  S_0[\varphi,{\tilde\varphi}]
  = \frac{1}{2}\int_{xy}\,
  \left( \varphi_a(x),{\tilde\varphi}_a(x) \right)
  \left(\begin{array}{cc}
  0         & iD^{-1}_{ab}(x,y)\\
  iD^{-1}_{ab}(x,y)    & 0
  \end{array}\right)
  \left(\begin{array}{c} \varphi_b(y) \\ {\tilde\varphi}_b(y) \end{array}\right)
  \label{eq:freeS}
\end{equation}
and the interaction part
\begin{equation}
  S_{\rm int}[\varphi,{\tilde\varphi}]
  = -\frac{\lambda}{6 N}\int_x {\tilde\varphi}_a(x)\varphi_a(x)\varphi_b(x)\varphi_b(x)
  \nonumber\\
  -\frac{\lambda}{24 N}\int_x {\tilde\varphi}_a(x){\tilde\varphi}_a(x)
                         {\tilde\varphi}_b(x)\varphi_b(x).
\label{eq:SqPhiXi}
\end{equation}
A similar construction can be done for the corresponding classical-statistical field theory, i.e.\ neglecting quantum corrections. For a detailed derivation see Refs.~\cite{Berges:2007ym,MSR}. The result for the classical-statistical generating functional is given by (\ref{eq:ZQuantPhiXi}) and (\ref{eq:freeS}) with $S_{\rm int}$ replaced by   
\begin{equation}
  S_{\rm int}^\mathrm{cl}[\varphi,{\tilde \varphi}] =
  -\frac{\lambda}{6 N}\int_x {\tilde \varphi}_a(x)\varphi_a(x)\varphi_b(x)\varphi_b(x) \, .
\label{eq:ClSqPhiXi}
\end{equation}
Comparing (\ref{eq:SqPhiXi}) and (\ref{eq:ClSqPhiXi}) one observes that the quantum theory includes the classical interaction vertex, however, in addition there is a vertex present which is absent classically.

From the generating functional for connected correlation functions
\begin{equation}
  W = -i\, {\rm ln} Z
  \label{eq:WGenF}
\end{equation}
we define the macroscopic field $\phi_a$, and ${\tilde\phi}_a$ by
\begin{equation}
  \frac{\delta W}{\delta {\tilde J}_a(x)} = \phi_a(x) \quad , \quad
  \frac{\delta W}{\delta J_a(x)}  = {\tilde\phi}_a(x)\, .
\label{eq:Defphiphitilde}
\end{equation}
The connected retarded/advanced propagators $G^\mathrm{R,A}_{ab}(x,y)$, the statistical correlation function $F_{ab}(x,y)$, and
${\tilde F}_{ab}(x,y)$ are defined by
\begin{eqnarray}
\frac{\delta^2 W}{\delta {\tilde J}_a(x) \delta J_b(y)}
& \!=\! &
 G^{\rm R}_{ab}(x,y) \quad , \quad
\frac{\delta^2 W}{\delta J_a(x) \delta {\tilde J}_b(y)}
\,=\, G^{\rm A}_{ab}(x,y) \, ,
\nonumber\\
\frac{\delta^2 W}{\delta {\tilde J}_a(x) \delta {\tilde J}_b(y)}
&\!=\!& 
i F_{ab}(x,y) \quad , \quad
\frac{\delta^2 W}{\delta J_a(x) \delta J_b(y)}
\,=\, i {\tilde F}_{ab}(x,y)\, ,
\label{eq:propdef}
\end{eqnarray}
and one observes 
\begin{equation}
G^\mathrm{A}_{ab}(x,y) = G^\mathrm{R}_{ba}(y,x) \, . 
\label{eq:GAGR}
\end{equation}
These are related to the spectral function defined in (\ref{eq:Frhodef}) by
\begin{equation}
\rho_{ab}(x,y) = G^\mathrm{R}_{ab}(x,y) - G^\mathrm{A}_{ab}(x,y) \, .
\label{eq:RGRGA}
\end{equation}
Similarly, we employ 
\begin{equation}
R^\mathrm{A}_{ab}(x,y) = R^\mathrm{R}_{ba}(y,x). 
\label{eq:RARR}
\end{equation}

We emphasize that the macroscopic fields and correlation functions depend on the sources. In particular, for vanishing sources $J$, ${\tilde J}$, $R^F$, $R^{\tilde F}$ the field ${\tilde\phi} = 0$, the function ${\tilde F} = 0$, and, more generally, $\delta^n W/\delta J(x_1) \delta J(x_2) \cdots \delta J(x_n)|_{J={\tilde J}=R^F=R^{\tilde F}=0}=0$ \cite{Chou:1984es}. It is important to note that this holds in the presence of non-vanishing retarded/advanced sources $R^\mathrm{R,A}$, which can be observed from the fact that these can be subsumed according to (\ref{eq:ZQuantPhiXi}) and (\ref{eq:freeS}) into a modified inverse classical propagator as $iD^{-1}+R^\mathrm{R,A}$. This will be used for the evaluation of the functional flow equation in Sec.~\ref{sec:exactflow}.

The Legendre transform of $W[J,\tilde{J},R^{\mathrm{R},\mathrm{A},F,{\tilde
  F}}]$ with respect to the source terms defines the two-particle irreducible ($2$PI) effective action~\cite{2PI,Chou:1984es,Calzetta:1986cq} 
\begin{eqnarray}
  \Gamma_{\rm 2PI}[\phi,{\tilde \phi},G^\mathrm{R},G^\mathrm{A},F,{\tilde F}]
  &\!=\!& W - \int_x \left(\phi_a(x) {\tilde J}_a(x)
     + {\tilde\phi}_a(x) J_a(x) \right)
  \nonumber\\
  &\!-\!&  \frac{1}{2} \int_{xy} \Big\{ R^\mathrm{A}_{ab}(x,y)
  \left(\phi_a(x){\tilde \phi}_b(y) - iG^\mathrm{R}_{ab}(x,y) \right)
  \nonumber\\
  &\!+\!&  R^\mathrm{R}_{ab}(x,y)
  \left({\tilde \phi}_a(x)\phi_b(y) -iG^\mathrm{A}_{ab}(x,y)\right)
  \nonumber\\
  &\!+\!&  R^{\tilde F}_{ab}(x,y)
    \Big(\phi_a(x)\phi_b(y) + F_{ab}(x,y) \Big)
  \nonumber\\
  &\!+\!&  R^{F}_{ab}(x,y)
  \left( {\tilde \phi}_a(x){\tilde \phi}_b(y) + {\tilde F}_{ab}(x,y) \right)
  \Big\}. 
  \label{eq:effact}
\end{eqnarray}
The $2$PI effective action has been extensively used to describe nonequilibrium dynamics. For a review and generalization to $n$PI functional integral techniques in this context see Ref.~\cite{Berges:2004yj}. A powerful nonperturbative approximation is the $2$PI $1/N$ expansion to NLO~\cite{Berges:2001fi,Aarts:2002dj}, which was successfully applied to far-from-equilibrium dynamics and late-time thermalization~\cite{Berges:2001fi,Berges:2002cz,Aarts:2001yn,Cooper:2002qd,Berges:2007ym,Berges:2008wm} as well as critical phenomena in thermal equilibrium~\cite{Alford:2004jj}. In Sec.~\ref{sec:2PIlargeN} we will recover the $2$PI $1/N$ expansion from the functional renormalization group (see also Ref.~\cite{Gasenzer:2007za}). The close connection between $n$PI and functional renormalization group techniques~\cite{Pawlowski:2005xe} will be exploited in the following for the discussion of nonthermal fixed points.

\subsection{Exact flow equation}
\label{sec:exactflow}

We consider the average action~\cite{Wetterich:1992yh}, evaluated along the closed time path as described above. The average action can be conveniently obtained from the 2PI effective action (\ref{eq:effact}) by partially undoing the Legendre transform with respect to the source terms quadratic in the fields:
\begin{equation}
\Gamma[\phi,{\tilde \phi};R^{\mathrm{R},\mathrm{A},F,{\tilde
  F}}] = \Gamma_{\rm 2PI} 
  - \frac{i}{2}\, {\rm Tr} \Big\{ G^\mathrm{R} R^\mathrm{R} 
  + G^\mathrm{A} R^\mathrm{A} 
  + i F R^{\tilde F}  
  + i {\tilde F} R^{F}  
  \Big\}.
  \label{eq:sourceaction}  
\end{equation}
Here we employ a compact notation, where the trace involves Minkowski space-time or momentum integration as well as summation over field indices. The remaining dependence on the bilinear source terms $R^{\mathrm{R},\mathrm{A},F,{\tilde F}}$ will be used in the 
following to construct the renormalization group flow of the one-particle irreducible
(1PI) effective action. The equations of motion for the fields are
\begin{eqnarray}
  \frac{\delta \Gamma[\phi,{\tilde \phi};R^{\mathrm{R},\mathrm{A},F,{\tilde
  F}}]}{\delta \phi_a(x)}
  \!&=&\! - {\tilde J}_a(x) -\int_y
      \Big( R^{\tilde F}_{ab}(x,y) \phi_b(y)
  \nonumber\\
  && +\ \frac{1}{2} R^\mathrm{A}_{ab}(x,y) {\tilde \phi}_b(y)
    +  \frac{1}{2}{\tilde \phi}_b(y) R^\mathrm{R}_{ba}(y,x) \Big),
  \nonumber\\
  \frac{\delta \Gamma[\phi,{\tilde \phi};R^{\mathrm{R},\mathrm{A},F,{\tilde
  F}}]}{\delta \tilde{\phi}_a(x)}
  \!&=&\! - J_a(x)
     -\int_y \Big( R^{F}_{ab}(x,y) \tilde{\phi}_b(y)
  \nonumber\\
  && +\ \frac{1}{2} R^\mathrm{R}_{ab}(x,y) \phi_b(y)
    +  \frac{1}{2} \phi_b(y) R^\mathrm{A}_{ba}(y,x) \Big),
    \label{eq:fields}
\end{eqnarray}
and  
\begin{eqnarray}
  \frac{\delta \Gamma[\phi,{\tilde \phi};R^{\mathrm{R},\mathrm{A},F,{\tilde
  F}}]}{\delta R^\mathrm{A}_{ab}(x,y)}
  &=& -\frac{i}{2} G^\mathrm{R}_{ab}(x,y)
  \quad , \quad
  \frac{\delta \Gamma[\phi,{\tilde \phi};R^{\mathrm{R},\mathrm{A},F,{\tilde
  F}}]}{\delta R^\mathrm{R}_{ab}(x,y)}
  \,=\, -\frac{i}{2} G^\mathrm{A}_{ab}(x,y) \, ,
  \nonumber\\
  \frac{\delta \Gamma[\phi,{\tilde \phi};R^{\mathrm{R},\mathrm{A},F,{\tilde
  F}}]}{\delta R^{\tilde F}_{ab}(x,y)}
  &=&
  \frac{1}{2} F_{ab}(x,y) \quad , \quad
  \frac{\delta \Gamma[\phi,{\tilde \phi};R^{\mathrm{R},\mathrm{A},F,{\tilde
  F}}]}{\delta R^{F}_{ab}(x,y)}
  \,=\, \frac{1}{2} {\tilde F}_{ab}(x,y) \, .
  \label{eq:derivR}
\end{eqnarray}  

As a consequence of the partial Legendre transform the quantity defined by  (\ref{eq:sourceaction}) becomes $R^{\mathrm{R},\mathrm{A},F,{\tilde
  F}}$-source dependent. For the renormalization group flow these source terms are taken to depend on some scale squared, $k^2 \ge 0$, which we denote as $R_k^{\mathrm{R},\mathrm{A},F,{\tilde
  F}}$. The average action 
\begin{equation}
\Gamma_k[\phi,{\tilde \phi}] \,=\, \Gamma[\phi,{\tilde \phi};R_k^{\mathrm{R},\mathrm{A},F,{\tilde
  F}}]
  \label{eq:averageaction}
\end{equation}  
as well as all correlation functions are taken to be scale dependent functionals of the fields $\phi$ and ${\tilde \phi}$, such that $F = F_k[\phi,{\tilde \phi}]$ etc. The behavior under changes of the scale is described by the $k^2$-derivative of (\ref{eq:sourceaction}). With the notation for the dimensionless "beta function" derivative $\dot{\Gamma}_k \equiv 2 k^2 \partial \Gamma_k/\partial k^2$ this yields
\begin{equation}
\dot{\Gamma}_k[\phi,{\tilde \phi}] = - \frac{i}{2} {\rm Tr} \Bigg\{  G^\mathrm{R}_k[\phi,{\tilde \phi}] \dot{R}_k^\mathrm{R}
  + G^\mathrm{A}_k[\phi,{\tilde \phi}] \dot{R}_k^\mathrm{A} 
  + i F_k[\phi,{\tilde \phi}]  \dot{R}_k^{\tilde F}   
  + i {\tilde F}_k[\phi,{\tilde \phi}] \dot{R}_k^{F}   
  \Bigg\}.
\label{eq:exactflow}  
\end{equation}   
This exact flow equation has a standard one-loop form, where the scale-dependent loop-propagators are related to second derivatives of the average action, e.g.\
\begin{equation}
\Gamma_{k,ab}^{\tilde{\phi}\phi}(x,y) \equiv \frac{\delta^2 \Gamma_k[\phi,\tilde{\phi}]}{\delta{\tilde \phi}_a(x) \delta\phi_b(y)} \, , \quad
\Gamma_{k,ab}^{\tilde{\phi}\tilde{\phi}}(x,y) \equiv \frac{\delta^2 \Gamma_k[\phi,{\tilde \phi}]}{\delta{\tilde \phi}_a(x) \delta{\tilde \phi}_b(y)} \, .
\end{equation}
In the compact matrix notation employed above these relations are given by the identities:
\begin{eqnarray}
G^\mathrm{R}_k[\phi,{\tilde \phi}] &\!=\!& -\left\{(\Gamma_k^{\tilde{\phi}\phi}+ R_k^\mathrm{R}) - (\Gamma_k^{\tilde{\phi}\tilde{\phi}}+R_k^F)(\Gamma_k^{\phi \tilde{\phi}}+R_k^\mathrm{A})^{-1}(\Gamma_k^{\phi \phi}+R_k^{\tilde{F}})\right\}^{-1}\!\! ,
\nonumber\\
G^\mathrm{A}_k[\phi,{\tilde \phi}] &\!=\!& -\left\{(\Gamma_k^{\phi\tilde{\phi}}+ R_k^\mathrm{A}) - (\Gamma_k^{\phi\phi}+R_k^{\tilde F})(\Gamma_k^{\tilde{\phi}\phi} + R_k^\mathrm{R})^{-1}(\Gamma_k^{\tilde{\phi}\tilde{\phi}}+R_k^F)\right\}^{-1}\!\! ,
\nonumber\\
i F_k[\phi,{\tilde \phi}] &\!=\!& -\left\{(\Gamma_k^{\phi\phi}+ R_k^{\tilde F}) - (\Gamma_k^{\phi\tilde{\phi}}+R_k^\mathrm{A})(\Gamma_k^{{\tilde \phi} \tilde{\phi}}+R_k^F)^{-1}(\Gamma_k^{{\tilde \phi} \phi}+R_k^\mathrm{R})\right\}^{-1}\!\! , 
\nonumber\\
i {\tilde F}_k[\phi,{\tilde \phi}] &\!=\!& -\left\{(\Gamma_k^{{\tilde \phi}{\tilde \phi}}+ R_k^F) - (\Gamma_k^{\tilde{\phi}\phi}+R_k^\mathrm{R})(\Gamma_k^{\phi\phi}+R_k^{\tilde F})^{-1}(\Gamma_k^{\phi{\tilde \phi}}+R_k^\mathrm{A})\right\}^{-1}\!\! , \quad
\label{eq:FG}
\end{eqnarray}
where we have used (\ref{eq:RARR}). These are obtained by noting that the relation between the fields and sources is $k$-dependent, i.e.~$J=J_k[\phi,{\tilde \phi}]$ and ${\tilde J}={\tilde J}_k[\phi,{\tilde \phi}]$, and a derivation is given in Appendix \ref{sec:funcform}.   

The sources $R_k^{\mathrm{R},\mathrm{A},F,{\tilde
  F}}$ are typically chosen from a family of functions to construct a renormalization group flow that interpolates between the classical action $S[\phi,\tilde{\phi}]$ for $k^2\to \infty$, or some UV cutoff scale $k^2 \to \Lambda^2$, and the standard 1PI effective action $\Gamma[\phi,\tilde{\phi}] = \Gamma[\phi,{\tilde \phi};R^{\mathrm{R},\mathrm{A},F,{\tilde
  F}} = 0]$ for $k^2\to 0$:
\begin{equation}
\lim\limits_{k^2 \to \Lambda^2} \Gamma_k[\phi,\tilde{\phi}] = S[\phi,\tilde{\phi}]\quad, \qquad
\lim\limits_{k^2 \to 0} \Gamma_k[\phi,\tilde{\phi}] = \Gamma[\phi,\tilde{\phi}]\,. 
\label{eq:interpolate}
\end{equation}  
This is standard for Euclidean field theories, however, on a closed time path there are few important differences which help to constrain the freedom in choosing $R_k^{\mathrm{R},\mathrm{A},F,{\tilde F}}$, which is discussed in the following.

Using the equations of motion for the fields (\ref{eq:fields}) with
\begin{eqnarray}
  \Gamma_{k,a}^{\phi}(x) \equiv \frac{\delta \Gamma_k[\phi,{\tilde \phi}]}{\delta \phi_a(x)}
  \quad , \quad
  \Gamma_{k,a}^{\tilde \phi}(x) \equiv \frac{\delta \Gamma_k[\phi,{\tilde \phi}]}{\delta \tilde{\phi}_a(x)} \, ,
    \label{eq:k-fields}
\end{eqnarray}
and the defining functional integral (\ref{eq:ZQuantPhiXi}) with (\ref{eq:WGenF}) and (\ref{eq:effact}) one can rewrite (\ref{eq:sourceaction}) as a functional integro-differential equation:\footnote{For the derivation one uses that the measures are invariant under the shifts
$\varphi \to \phi+\varphi$ and ${\tilde \varphi} \to {\tilde \phi}
+ {\tilde \varphi}$.}    
\begin{eqnarray}
\Gamma_k[\phi,{\tilde \phi}] \!&=&\! S[\phi,{\tilde \phi}] -i \ln \int {\cal D}\varphi{\cal D} {\tilde\varphi}
     \,\exp\Bigg\{i\Bigg[S[\phi+\varphi,{\tilde \phi}+{\tilde\varphi}]
-S[\phi,{\tilde \phi}]     \nonumber\\
&-& \! \int_x\,\left( \varphi_a(x) \Gamma_{k,a}^{\phi}(x) + {\tilde\varphi}_a(x) \Gamma_{k,a}^{\tilde \phi}(x) \right)
  \nonumber\\
  &+& \! \frac{1}{2}\int_{xy}\!\!
  \left( \varphi_a(x),{\tilde\varphi}_a(x) \right)
  \left(\!\begin{array}{cc}
  R^{\tilde F}_{k,ab}(x,y)        \! &\! R^\mathrm{A}_{k,ab}(x,y)\\[1pt]
  R^\mathrm{R}_{k,ab}(x,y)        \! &\! R^{F}_{k,ab}(x,y)
  \end{array}\!\right)
  \left(\!\!\begin{array}{c} \varphi_b(y) \\ {\tilde\varphi}_b(y)
  \end{array}\!\!\right)
  \Bigg]\Bigg\}.\quad
  \label{eq:integrodiff}
\end{eqnarray}
With the representation of the delta functional
\begin{equation}
\int {\cal D}{\tilde \varphi} \exp \Bigg\{i \int_x
\varphi_a(x) {\tilde\varphi}_a(x)\Bigg\}
\, = \, \delta \left[ \varphi \right]
\label{eq:delta}
\end{equation}
one observes from (\ref{eq:integrodiff}) that the interpolation property (\ref{eq:interpolate}) can be efficiently implemented with the class of functions 
\begin{equation}
R^{\mathrm R,A}_{k,ab}(x,y) = R_k(-\square_x) \delta_{ab} \delta(x-y)
\quad, \quad R^{F,{\tilde F}}_{k,ab}(x,y) = 0\, ,
\label{eq:Rchoice}
\end{equation}
where in Fourier space for fixed momentum $p^2$ the infrared cutoff function is required to behave as 
\begin{equation}
\lim\limits_{k^2 \to \Lambda^2} R_k(p^2) \to \infty \quad, \qquad
\lim\limits_{k^2 \to 0} R_k(p^2) \to 0\,. 
\label{eq:limitsR}
\end{equation} 
We emphasize that $R^{F,\tilde F}_k$ cannot be chosen to diverge as $R^{\mathrm R,A}_k$ for large $k^2$ since the latter has to dominate for short-distance scales in order to recover the $\delta$-function behavior (\ref{eq:delta}). For $R^{F,\tilde F}_k = 0$ one obtains from (\ref{eq:exactflow}) 
\begin{equation}
\dot{\Gamma}_k[\phi,{\tilde \phi}] = - \frac{i}{2}{\rm Tr} \left\{    
G^\mathrm{R}_k[\phi,{\tilde \phi}] \dot{R}_k^\mathrm{R} + G^\mathrm{A}_k[\phi,{\tilde \phi}] \dot{R}_k^\mathrm{A}
\right\}.
\label{eq:exactflow2}  
\end{equation}
This exact flow equation will be used below for the description of nonthermal fixed points. 
With (\ref{eq:GAGR}) and (\ref{eq:RARR}) it can be written in a more compact form using
${\rm Tr} \left\{ G^\mathrm{R}_k[\phi,{\tilde \phi}] \dot{R}_k^\mathrm{R} \right\} = {\rm Tr} \left\{ G^\mathrm{A}_k[\phi,{\tilde \phi}] \dot{R}_k^\mathrm{A} \right\}$. However, we will keep the explicit symmetry between retarded and advanced functions in the notation to simplify the diagrammatics below. The fact that in (\ref{eq:exactflow2}) the "off-diagonal" retarded/advanced functions $R_k^{\mathrm R,A}$ are sufficient to implement an infrared cutoff on fluctuations coincides with the property that in the defining functional integral (\ref{eq:ZQuantPhiXi}) the free propagator (\ref{eq:freeS}) exhibits vanishing "diagonal" entries. This motivates similar constructions of flow equations by modifying the frequency or momentum dependence of free propagators as in Refs.~\cite{Gezzi,Jakobs}. 

Exact flow equations for $n$-point functions are obtained from $n$ functional derivatives of (\ref{eq:exactflow2}) with respect to the fields $\phi$ and ${\tilde \phi}$. For instance,  
\begin{equation}
\dot{\Gamma}_{k,ab}^{{\tilde \phi}{\tilde \phi}}(x,y)= - \frac{i}{2}{\rm Tr} \left\{    
\frac{\delta^2 G^\mathrm{R}_k[\phi,{\tilde \phi}]}{\delta {\tilde \phi}_a(x) \delta {\tilde \phi}_b(y)} \dot{R}_k^\mathrm{R} + \frac{\delta^2 G^\mathrm{A}_k[\phi,{\tilde \phi}]}{\delta {\tilde \phi}_a(x) \delta {\tilde \phi}_b(y)} \dot{R}_k^\mathrm{A}
\right\},
\label{eq:exactflowtptp}  
\end{equation}
where the derivatives will be evaluated in Sec.~\ref{sec:dia} using (\ref{eq:FG}). Since the latter equations for $G^{\mathrm R,A}_k[\phi,{\tilde \phi}]$ contain already second derivatives of $\Gamma_k[\phi,{\tilde \phi}]$, one recovers the fact that flow equations for $n$-point functions depend on up to $(n+2)$-point functions.

\subsection{Diagrammatics}
\label{sec:dia}

After taking derivatives of (\ref{eq:exactflow2}) to obtain\footnote{The diagrammatic results derived from (\ref{eq:exactflow2}) or from (\ref{eq:exactflow}) agree, of course.} exact flow equations for $n$-point functions, all sources can be set to zero except for $R_k^{\mathrm R,A}$, which implement the scale dependence as discussed above. Following Sec.~\ref{sec:genfunc} this corresponds to ${\tilde \phi} = 0$ and ${\tilde F}_k = 0$, while the latter entails $\Gamma_k^{\phi\phi}=0$ according to (\ref{eq:FG}). More generally, $\delta^n \Gamma[\phi,{\tilde \phi}]/\delta \phi(x_1) \delta \phi(x_2) \cdots \delta \phi(x_n)|_{\phi=\phi_0,{\tilde \phi}=0}=0$, where the macroscopic field in the absence of sources, $\phi = \phi_0$, may be nonzero corresponding to spontaneous symmetry breaking~\cite{Chou:1984es}. From (\ref{eq:FG}) one finds
\begin{eqnarray}
&&  \begin{picture}(370,21) (38,-206)
    \SetWidth{0.5}
    \SetColor{Black}
    \Text(40,-206)[lb]{$\displaystyle G_{k,ab}^{\mathrm R}[\phi = \phi_0,{\tilde \phi} = 0] \,=\, - \left( \Gamma_k^{\tilde{\phi}\phi} + R_k^{\mathrm R} \right)^{-1}_{ab} [\phi = \phi_0,{\tilde \phi} = 0] \, = \, $}
   \Text(344,-193)[lb]{$a$}
   \Text(371,-193)[lb]{$b$}
\Text(382,-199)[lb]{,}
    \SetWidth{1.0}
    \Line(342,-195)(356,-195)
    \DashLine(356,-195)(376,-195){2}
  \end{picture}
\nonumber\\
&&  \begin{picture}(370,23) (38,-202)
    \SetWidth{0.5}
    \SetColor{Black}
    \Text(40,-206)[lb]{$\displaystyle G_{k,ab}^{\mathrm A}[\phi = \phi_0,{\tilde \phi} = 0] \,=\, - \left( \Gamma_k^{\phi\tilde{\phi}} + R_k^{\mathrm A} \right)^{-1}_{ab} [\phi = \phi_0,{\tilde \phi} = 0] \, = \, $}
   \Text(344,-193)[lb]{$a$}
   \Text(371,-193)[lb]{$b$}
   \Text(382,-199)[lb]{,}
    \SetWidth{1.0}
    \DashLine(342,-195)(356,-195){2}
    \Line(358,-195)(376,-195)
  \end{picture}
\nonumber\\  
&&  \begin{picture}(370,23) (38,-202)
    \SetWidth{0.5}
    \SetColor{Black}
    \Text(40,-206)[lb]{$\displaystyle  iF_{k,ab}[\phi = \phi_0,{\tilde \phi} = 0] \,=\,\,\, 
    \left( G_{k}^{\mathrm R}\, \Gamma^{\tilde{\phi}\tilde{\phi}}_{k}\, G_{k}^{\mathrm A}\right)_{ab} [\phi = \phi_0,{\tilde \phi} = 0] \,\, = \, $}
   \Text(344,-193)[lb]{$a$}
   \Text(371,-193)[lb]{$b$}
   \Text(382,-199)[lb]{.}
    \SetWidth{1.0}
    \Line(342,-195)(376,-195)
  \end{picture}
\label{eq:dia}
\end{eqnarray}
For a diagrammatic description a solid line is associated to the statistical correlation function $iF_k$, and the retarded (advanced) propagator $G_k^{\mathrm R}$ ($G_k^{\mathrm A}$) is represented by a solid-dashed (dashed-solid) line. The cutoff function derivatives $\dot{R}_k^{\mathrm R}$ and $\dot{R}_k^{\mathrm A}$ are associated to a cross, with reversed ordering of dashed to solid lines as compared to the retarded and advanced propagators: 
\begin{equation}
\begin{picture}(320,-185) (28,-192)
    \SetWidth{0.5}
    \SetColor{Black}
    \Text(72,-198)[lb]{$\dot{R}_{k,ab}^{\mathrm R} \,=\, $}
    \Text(212,-198)[lb]{$\dot{R}_{k,ab}^{\mathrm A} \,=\, $}
\Text(125,-188)[lb]{$a$}
\Text(170,-188)[lb]{$b$}
\Text(265,-188)[lb]{$a$}
\Text(310,-188)[lb]{$b$}
\Text(178,-195)[lb]{\quad,}
\Text(325,-192)[lb]{.}
    \SetWidth{1.0}
    \Line(145,-193)(151,-187)\Line(145,-187)(151,-193)
    \DashLine(151,-190)(124,-190){2}
    \Line(151,-190)(175,-190)
\SetWidth{1.0}
    \Line(288,-193)(294,-187)\Line(288,-187)(294,-193)
    \Line(291,-190)(264,-190)
    \DashLine(291,-190)(315,-190){2}
\end{picture}
\label{eq:RRRAdia}
\end{equation}  
Similarly, the proper vertices obtained from $n$ functional $\tilde \phi$-derivatives and 
$m$ functional $\phi$-derivatives of $\Gamma_k[\phi,{\tilde \phi}]$ are represented by full dots connected to $n$ dashed lines and $m$ solid lines, once the fields are evaluated at $\phi = \phi_0$ and ${\tilde \phi} = 0$. Four-point vertices are displayed, for instance, as 
\begin{equation}
\begin{picture}(20,29) (293,-158)
    \SetWidth{0.5}
    \SetColor{Black}
\Text(190,-153)[lb]{$\Gamma_{k,abcd}^{{\tilde \phi}\phi\phi\phi} \,=\, $}
\Text(244,-135)[lb]{$c$}
\Text(273,-135)[lb]{$d$}
\Text(243,-161)[lb]{$a$}
\Text(273,-161)[lb]{$b$}
\Text(275,-152)[lb]{\quad,}
    \Text(325,-153)[lb]{$\Gamma_{k,abcd}^{{\tilde \phi}{\tilde \phi}\phi\phi} \,=\, $}
    \Text(375,-135)[lb]{$c$}
    \Text(403,-135)[lb]{$d$}
    \Text(374,-161)[lb]{$a$}
    \Text(403,-161)[lb]{$b$}
    \Text(410,-150)[lb]{.}
\SetWidth{1.0}
    \Vertex(260,-145){3.4}
    \Line(250,-135)(270,-155)
    \Line(260,-145)(270,-135)
    \DashLine(260,-145)(250,-155){2}
\SetWidth{1.0}
    \Vertex(390,-145){3.4}
    \Line(380,-135)(390,-145)
    \Line(390,-146)(400,-135)
    \DashLine(390,-145)(380,-155){2}
    \DashLine(400,-155)(390,-145){2}
\end{picture}
\end{equation}
We will frequently denote proper four-point vertices as $\Gamma^{(4)}_{k,abcd}$, irrespective of their external lines, and similarly for two-point functions $\Gamma^{(2)}_{k,ab}$ etc. 

With this notation the exact renormalization group equation (\ref{eq:exactflow2}) reads for $\dot{\Gamma}_k = \dot{\Gamma}_k [\phi=\phi_0,\tilde{\phi}=0]$ diagrammatically:
\begin{equation}
  \begin{picture}(93,23) (25,-165)
    \SetWidth{0.5}
    \SetColor{Black}
    \Text(-10,-167)[lb]{$\dot{\Gamma}_k \,=\, -\frac{i}{2} \Big\{ $}
    \Text(80,-160)[lb]{$\displaystyle +$}
    \Text(126,-167)[lb]{$\Big\}$ .}
    \SetWidth{1.0}
    \Line(58,-148)(64,-142)\Line(58,-142)(64,-148)
    \CArc(61,-155)(10,90,270)
    \DashCArc(61,-155)(10,-90,90){2}
\SetWidth{1.0}
    \Line(108,-148)(114,-142)\Line(108,-142)(114,-148)
    \DashCArc(111,-155)(10,90,270){2}
    \CArc(111,-155)(10,-90,90)
  \end{picture}
\label{eq:exactGdia}  
\end{equation}
It should be emphasized that diagrammatic descriptions of flow equations employ (\ref{eq:dia}), which exploits the vanishing of the anomalous propagator ${\tilde F}_k$ and $\Gamma_k^{\phi\phi}$ at extrema of the generating functional. In contrast, the renormalization group equation (\ref{eq:exactflow2}) using the full functional form of the propagators (\ref{eq:FG}) is still required to obtain the flow of $n$-point correlation functions by further functional differentiation. 

In view of the application to nonthermal fixed points, we will consider in the following no spontaneous symmetry breaking ($\phi_0 =0$) for simplicity, though the generalization to $\phi_0 \not =0$ is straightforward. In the absence of spontaneous symmetry breaking, where three-point vertices vanish for the theory described by (\ref{eq:Sq}), we derive the diagrammatic expressions for $\dot{\Gamma}_{k,ab}^{(2)}$ and $\dot{\Gamma}_{k,abcd}^{(4)}$ from derivatives of (\ref{eq:exactflow2}) employing the functional form of the propagators (\ref{eq:FG}). This is used to observe standard diagrammatic rules, which facilitate the computation of flow equations for higher $n$-point correlation functions. 

According to (\ref{eq:exactflowtptp}) the exact flow equation for the two-point function $\Gamma_k^{{\tilde \phi}{\tilde \phi}}$ involves the functional derivative 
\begin{eqnarray}
\frac{\delta^2 G^\mathrm{R}_{k,ab}[\phi,{\tilde \phi}]}{\delta {\tilde \phi}_c \delta {\tilde \phi}_d} 
&\!=\!& \left\{ G^{\mathrm R}_{k,ae} \Gamma^{{\tilde \phi}\phi{\tilde \phi}{\tilde \phi}}_{k,efcd} G^{\mathrm R}_{k,fb}
- G^{\mathrm R}_{k,ae} \Gamma^{{\tilde \phi}{\tilde \phi}{\tilde \phi}{\tilde \phi}}_{k,efcd} \left( \Gamma^{\phi{\tilde \phi}} +R^{\mathrm A}\right)^{-1}_{k,fg} \Gamma^{\phi \phi}_{k,gh} G^{\mathrm R}_{k,hb} \right. \nonumber\\
&\!+\!& G^{\mathrm R}_{k,ae} \Gamma^{{\tilde \phi}{\tilde \phi}}_{k,ef} \left( \Gamma^{\phi{\tilde \phi}} +R^{\mathrm A}\right)^{-1}_{k,fg} \Gamma^{\phi{\tilde \phi}{\tilde \phi}{\tilde \phi}}_{k,ghcd} \left( \Gamma^{\phi{\tilde \phi}} +R^{\mathrm A}\right)^{-1}_{k,hi} \Gamma^{\phi \phi}_{k,ij} G^{\mathrm R}_{k,jb} \nonumber\\
&\!-\!& \left. G^{\mathrm R}_{k,ae} \Gamma^{{\tilde \phi}{\tilde \phi}}_{k,ef} \left( \Gamma^{\phi{\tilde \phi}} +R^{\mathrm A}\right)^{-1}_{k,fg} \Gamma^{\phi \phi {\tilde \phi}{\tilde \phi}}_{k,ghcd} G^{\mathrm R}_{k,hb}
\right\}[\phi,{\tilde \phi}] \, ,
\label{eq:derivG}
\end{eqnarray}
suppressing space-time arguments and neglecting three-point vertices. This is obtained from (\ref{eq:FG}) with $R^{F,\tilde F}_k = 0$. Since $\Gamma_k^{\phi\phi}[\phi=0,\tilde{\phi}=0]=0$, only the first and last term of (\ref{eq:derivG}) contribute to the compact diagrammatic expression for $\dot{\Gamma}_{k,ab}^{{\tilde \phi}{\tilde \phi}}$. Accordingly, for $\dot{\Gamma}_{k}^{(2)} = \dot{\Gamma}_k^{(2)} [\phi=0,\tilde{\phi}=0]$ we find
\begin{equation}
  \begin{picture}(349,34) (-23,-52)
    \SetWidth{0.5}
    \SetColor{Black}
    \Text(140,-39)[lb]{$\displaystyle +$}
    \Text(190,-39)[lb]{$\displaystyle +$}
    \Text(240,-39)[lb]{$\displaystyle +$}
    \Text(290,-48)[lb]{$\Big\}\, .$}
    \Text(30,-48)[lb]{$\displaystyle 
    \dot{\Gamma}_{k,ab}^{(2)} 
    \, =\, -\frac{i}{2} \, \Big\{$}
  \Text(110,-55)[lb]{$a$}
  \Text(126,-55)[lb]{$b$}
  \Text(160,-55)[lb]{$a$}
  \Text(176,-55)[lb]{$b$}
  \Text(210,-55)[lb]{$a$}
  \Text(226,-55)[lb]{$b$}
  \Text(260,-55)[lb]{$a$}
  \Text(276,-55)[lb]{$b$}
    \SetWidth{1.0}
    \CArc(120,-34)(10,100,180)
    \DashCArc(120,-34)(10,180,-80){2}
    \DashCArc(120,-34)(10,0,100){2}
    \CArc(120,-34)(10,-80,0)
    \SetWidth{0.5}
    \Vertex(120,-44){3.4}
    \SetWidth{1.0}
    \Line(117,-21)(123,-27)\Line(117,-27)(123,-21)
    \Line(110,-45)(130,-45)
    \CArc(170,-34)(10,90,270)
    \DashCArc(170,-34)(10,0,90){2}
    \CArc(170,-34)(10,-90,0)
   \SetWidth{0.5}
    \Vertex(170,-44){3.4}
    \SetWidth{1.0}
    \Line(167,-21)(173,-27)\Line(167,-27)(173,-21)
    \Line(160,-45)(180,-45)
   \SetWidth{1.0}
    \DashCArc(220,-34)(10,90,180){2}
    \CArc(220,-34)(10,180,270)
    \CArc(220,-34)(10,0,90)
    \DashCArc(220,-34)(10,-90,0){2}
    \SetWidth{0.5}
    \Vertex(220,-44){3.4}
    \SetWidth{1.0}
    \Line(217,-21)(223,-27)\Line(217,-27)(223,-21)
    \Line(210,-45)(230,-45)
    \CArc(270,-34)(10,180,270)
    \DashCArc(270,-34)(10,90,180){2}
    \CArc(270,-34)(10,-90,90)
    \SetWidth{0.5}
    \Vertex(270,-44){3.4}
    \SetWidth{1.0}
    \Line(267,-21)(273,-27)\Line(267,-27)(273,-21)
    \Line(260,-45)(280,-45)
  \end{picture}
\label{eq:exactG2dia}    
\end{equation}
Since a four-point function involves two further functional derivatives, it receives also contributions from derivatives of the anomalous terms $\Gamma_k^{\phi\phi}[\phi,\tilde{\phi}]$. Evaluated at $\phi=\tilde{\phi}=0$ these do not vanish in general, since only $\Gamma_k^{\phi\phi\phi\phi}[\phi=0,\tilde{\phi}=0]=0$. We obtain
\begin{eqnarray}  
&&
\begin{picture}(349,58) (35,-75)
    \SetWidth{0.5}
    \SetColor{Black}
    \Text(130,-42)[lb]{$\displaystyle +$}
    \Text(180,-42)[lb]{$\displaystyle +$}
    \Text(230,-42)[lb]{$\displaystyle +$}
    \Text(280,-42)[lb]{$\displaystyle +$}
    \Text(330,-42)[lb]{$\displaystyle +$}
    \Text(12,-50)[lb]{$\displaystyle 
    \dot{\Gamma}_{k,abcd}^{(4)} 
    \, =\, -\frac{i}{8} \, \Big\{$}
    \Text(90,-60)[lb]{$a$}
    \Text(102,-60)[lb]{$b$}
    \Text(112,-60)[lb]{$c$}
    \Text(122,-60)[lb]{$d$}
    \Text(140,-60)[lb]{$a$}
    \Text(152,-60)[lb]{$b$}
    \Text(162,-60)[lb]{$c$}
    \Text(172,-60)[lb]{$d$}
    \Text(190,-60)[lb]{$a$}
    \Text(202,-60)[lb]{$b$}
    \Text(212,-60)[lb]{$c$}
    \Text(222,-60)[lb]{$d$}
    \Text(240,-60)[lb]{$a$}
    \Text(252,-60)[lb]{$b$}
    \Text(262,-60)[lb]{$c$}
    \Text(272,-60)[lb]{$d$}
    \Text(290,-60)[lb]{$a$}
    \Text(302,-60)[lb]{$b$}
    \Text(312,-60)[lb]{$c$}
    \Text(322,-60)[lb]{$d$}
    \Text(340,-60)[lb]{$a$}
    \Text(352,-60)[lb]{$b$}
    \Text(362,-60)[lb]{$c$}
    \Text(372,-60)[lb]{$d$}
    \SetWidth{1.0}
    \CArc(110,-34)(10,90,140)
    \DashCArc(110,-34)(10,140,-80){2}
    \DashCArc(110,-34)(10,20,90){2}
    \CArc(110,-34)(10,-80,20)
    \SetWidth{0.5}
    \Vertex(102,-39){3.4}
    \Vertex(118,-39){3.4}
    \SetWidth{1.0}
    \Line(107,-21)(113,-27)\Line(107,-27)(113,-21)
    \Line(102,-39)(95,-51)
    \Line(102,-39)(105,-51)
    \Line(118,-39)(115,-51)
    \Line(118,-39)(125,-51)
    \CArc(160,-34)(10,90,140)
    \DashCArc(160,-34)(10,20,90){2}
    \DashCArc(160,-34)(10,140,210){2}
    \CArc(160,-34)(10,-30,20)
    \CArc(160,-34)(10,210,260)
    \DashCArc(160,-34)(10,260,-30){2}
    \SetWidth{0.5}
    \Vertex(152,-39){3.4}
    \Vertex(168,-39){3.4}
    \SetWidth{1.0}
    \Line(157,-21)(163,-27)\Line(157,-27)(163,-21)
    \Line(152,-39)(145,-51)
    \Line(152,-39)(155,-51)
    \Line(168,-39)(165,-51)
    \Line(168,-39)(175,-51)
    \CArc(210,-34)(10,90,140)
    \DashCArc(210,-34)(10,20,90){2}
    \DashCArc(210,-34)(10,140,210){2}
    \CArc(210,-34)(10,210,20)
    \SetWidth{0.5}
    \Vertex(202,-39){3.4}
    \Vertex(218,-39){3.4}
    \SetWidth{1.0}
    \Line(207,-21)(213,-27)\Line(207,-27)(213,-21)
    \Line(202,-39)(195,-51)
    \Line(202,-39)(205,-51)
    \Line(218,-39)(215,-51)
    \Line(218,-39)(225,-51)
\SetWidth{1.0}
    \CArc(260,-34)(10,90,210)
    \DashCArc(260,-34)(10,210,280){2}
    \DashCArc(260,-34)(10,20,90){2}
    \CArc(260,-34)(10,-80,20)
    \SetWidth{0.5}
   \Vertex(252,-39){3.4}
    \Vertex(268,-39){3.4}
    \SetWidth{1.0}
    \Line(257,-21)(263,-27)\Line(257,-27)(263,-21)
    \Line(252,-39)(245,-51)
    \Line(252,-39)(255,-51)
    \Line(268,-39)(265,-51)
    \Line(268,-39)(275,-51)
    \CArc(310,-34)(10,90,260)
    \DashCArc(310,-34)(10,20,90){2}
    \DashCArc(310,-34)(10,260,-30){2}
    \CArc(310,-34)(10,-30,20)
    \SetWidth{0.5}
   \Vertex(302,-39){3.4}
    \Vertex(318,-39){3.4}
    \SetWidth{1.0}
    \Line(307,-21)(313,-27)\Line(307,-27)(313,-21)
    \Line(302,-39)(295,-51)
    \Line(302,-39)(305,-51)
    \Line(318,-39)(315,-51)
    \Line(318,-39)(325,-51)
    \CArc(360,-34)(10,90,20)
    \DashCArc(360,-34)(10,20,90){2}
    \Vertex(352,-39){3.4}
    \Vertex(368,-39){3.4}
    \SetWidth{1.0}
    \Line(357,-21)(363,-27)\Line(357,-27)(363,-21)
    \Line(352,-39)(345,-51)
    \Line(352,-39)(355,-51)
    \Line(368,-39)(365,-51)
    \Line(368,-39)(375,-51)
  \end{picture}
\nonumber\\
&&  
\begin{picture}(349,48) (35,-75)
    \SetWidth{0.5}
    \SetColor{Black}
    \Text(130,-42)[lb]{$\displaystyle +$}
    \Text(180,-42)[lb]{$\displaystyle +$}
    \Text(230,-42)[lb]{$\displaystyle +$}
    \Text(280,-42)[lb]{$\displaystyle +$}
    \Text(330,-42)[lb]{$\displaystyle +$}
    \Text(80,-42)[lb]{$\displaystyle +$}
    \Text(90,-60)[lb]{$a$}
    \Text(102,-60)[lb]{$b$}
    \Text(112,-60)[lb]{$c$}
    \Text(122,-60)[lb]{$d$}
    \Text(140,-60)[lb]{$a$}
    \Text(152,-60)[lb]{$b$}
    \Text(162,-60)[lb]{$c$}
    \Text(172,-60)[lb]{$d$}
    \Text(190,-60)[lb]{$a$}
    \Text(202,-60)[lb]{$b$}
    \Text(212,-60)[lb]{$c$}
    \Text(222,-60)[lb]{$d$}
    \Text(240,-60)[lb]{$a$}
    \Text(252,-60)[lb]{$b$}
    \Text(262,-60)[lb]{$c$}
    \Text(272,-60)[lb]{$d$}
    \Text(290,-60)[lb]{$a$}
    \Text(302,-60)[lb]{$b$}
    \Text(312,-60)[lb]{$c$}
    \Text(322,-60)[lb]{$d$}
    \Text(340,-60)[lb]{$a$}
    \Text(352,-60)[lb]{$b$}
    \Text(362,-60)[lb]{$c$}
    \Text(372,-60)[lb]{$d$}
    \SetWidth{1.0}
\CArc(110,-34)(10,40,90)
    \DashCArc(110,-34)(10,-30,40){2}
    \DashCArc(110,-34)(10,90,160){2}
    \CArc(110,-34)(10,160,210)
    \DashCArc(110,-34)(10,210,280){2}
    \CArc(110,-34)(10,-80,-30)
    \SetWidth{0.5}
    \Vertex(102,-39){3.4}
    \Vertex(118,-39){3.4}
    \SetWidth{1.0}
    \Line(107,-21)(113,-27)\Line(107,-27)(113,-21)
    \Line(102,-39)(95,-51)
    \Line(102,-39)(105,-51)
    \Line(118,-39)(115,-51)
    \Line(118,-39)(125,-51)
    \CArc(160,-34)(10,40,90)
    \DashCArc(160,-34)(10,260,40){2}
    \CArc(160,-34)(10,160,260)
    \DashCArc(160,-34)(10,90,160){2}
    \SetWidth{0.5}
    \Vertex(152,-39){3.4}
    \Vertex(168,-39){3.4}
    \SetWidth{1.0}
    \Line(157,-21)(163,-27)\Line(157,-27)(163,-21)
    \Line(152,-39)(145,-51)
    \Line(152,-39)(155,-51)
    \Line(168,-39)(165,-51)
    \Line(168,-39)(175,-51)
    \CArc(210,-34)(10,40,90)
    \DashCArc(210,-34)(10,-30,40){2}
    \CArc(210,-34)(10,160,-30)
    \DashCArc(210,-34)(10,90,160){2}
    \SetWidth{0.5}
    \Vertex(202,-39){3.4}
    \Vertex(218,-39){3.4}
    \SetWidth{1.0}
    \Line(207,-21)(213,-27)\Line(207,-27)(213,-21)
    \Line(202,-39)(195,-51)
    \Line(202,-39)(205,-51)
    \Line(218,-39)(215,-51)
    \Line(218,-39)(225,-51)
\SetWidth{1.0}
    \CArc(260,-34)(10,-80,90)
    \DashCArc(260,-34)(10,90,160){2}
    \CArc(260,-34)(10,160,210)
    \DashCArc(260,-34)(10,210,280){2}
    \SetWidth{0.5}
   \Vertex(252,-39){3.4}
    \Vertex(268,-39){3.4}
    \SetWidth{1.0}
    \Line(257,-21)(263,-27)\Line(257,-27)(263,-21)
    \Line(252,-39)(245,-51)
    \Line(252,-39)(255,-51)
    \Line(268,-39)(265,-51)
    \Line(268,-39)(275,-51)
    \CArc(310,-34)(10,-30,90)
    \DashCArc(310,-34)(10,260,-30){2}
    \CArc(310,-34)(10,160,260)
    \DashCArc(310,-34)(10,90,160){2}
    \SetWidth{0.5}
   \Vertex(302,-39){3.4}
    \Vertex(318,-39){3.4}
    \SetWidth{1.0}
    \Line(307,-21)(313,-27)\Line(307,-27)(313,-21)
    \Line(302,-39)(295,-51)
    \Line(302,-39)(305,-51)
    \Line(318,-39)(315,-51)
    \Line(318,-39)(325,-51)
    \CArc(360,-34)(10,160,90)
    \DashCArc(360,-34)(10,90,160){2}
    \Vertex(352,-39){3.4}
    \Vertex(368,-39){3.4}
    \SetWidth{1.0}
    \Line(357,-21)(363,-27)\Line(357,-27)(363,-21)
    \Line(352,-39)(345,-51)
    \Line(352,-39)(355,-51)
    \Line(368,-39)(365,-51)
    \Line(368,-39)(375,-51)
  \end{picture}
\nonumber\\
&&  
\begin{picture}(349,28) (35,-60)
    \SetWidth{0.5}
    \SetColor{Black}
 \Text(80,-42)[lb]{$\displaystyle + \quad P(a,b,c,d) \, \Big\}$}
\end{picture}
 \nonumber\\
&&
\begin{picture}(349,26) (35,-64)
    \SetWidth{0.5}
    \SetColor{Black}
    \Text(130,-44)[lb]{$\displaystyle +$}
    \Text(180,-44)[lb]{$\displaystyle +$}
    \Text(230,-45)[lb]{$\displaystyle +$}
    \Text(280,-47)[lb]{$\Big\} \, ,$}
    \Text(62,-47)[lb]{$\displaystyle - \frac{i}{2}\, \Big\{$}
\Text(93,-63)[lb]{$a$}
\Text(102,-63)[lb]{$b$}
\Text(113,-63)[lb]{$c$}
\Text(122,-63)[lb]{$d$}
\Text(143,-63)[lb]{$a$}
\Text(152,-63)[lb]{$b$}
\Text(163,-63)[lb]{$c$}
\Text(172,-63)[lb]{$d$}
\Text(193,-63)[lb]{$a$}
\Text(202,-63)[lb]{$b$}
\Text(213,-63)[lb]{$c$}
\Text(222,-63)[lb]{$d$}
\Text(243,-63)[lb]{$a$}
\Text(252,-63)[lb]{$b$}
\Text(263,-63)[lb]{$c$}
\Text(272,-63)[lb]{$d$}
\SetWidth{1.0}
    \CArc(110,-34)(10,90,180)
    \DashCArc(110,-34)(10,180,270){2}
    \DashCArc(110,-34)(10,0,90){2}
    \CArc(110,-34)(10,-90,0)
    \SetWidth{0.5}
    \Vertex(110,-44){3.4}
    \SetWidth{1.0}
    \Line(107,-21)(113,-27)\Line(107,-27)(113,-21)
    \Line(97,-55)(110,-44)
    \Line(105,-55)(110,-44)
    \Line(115,-55)(110,-44)
    \Line(123,-55)(110,-44)
    \CArc(160,-34)(10,90,270)
    \DashCArc(160,-34)(10,0,90){2}
    \CArc(160,-34)(10,-90,0)
   \SetWidth{0.5}
    \Vertex(160,-44){3.4}
    \SetWidth{1.0}
    \Line(157,-21)(163,-27)\Line(157,-27)(163,-21)
    \Line(147,-55)(160,-44)
    \Line(155,-55)(160,-44)
    \Line(165,-55)(160,-44)
    \Line(173,-55)(160,-44)
    \DashCArc(210,-34)(10,90,180){2}
    \CArc(210,-34)(10,180,270)
    \CArc(210,-34)(10,0,90)
    \DashCArc(210,-34)(10,-90,0){2}
    \SetWidth{0.5}
    \Vertex(210,-44){3.4}
    \SetWidth{1.0}
    \Line(207,-21)(213,-27)\Line(207,-27)(213,-21)
    \Line(197,-55)(210,-44)
    \Line(205,-55)(210,-44)
    \Line(215,-55)(210,-44)
    \Line(223,-55)(210,-44)
    \CArc(260,-34)(10,180,90)
    \DashCArc(260,-34)(10,90,180){2}
   \SetWidth{0.5}
    \Vertex(260,-44){3.4}
    \SetWidth{1.0}
    \Line(257,-21)(263,-27)\Line(257,-27)(263,-21)
    \Line(247,-55)(260,-44)
    \Line(255,-55)(260,-44)
    \Line(265,-55)(260,-44)
    \Line(273,-55)(260,-44)
  \end{picture}
\label{eq:exactG4dia}    
\end{eqnarray}
where $P(a,b,c,d)$ denotes permutations of the respective field indices.
Likewise, flow equations for six-point or higher vertices can be derived by further functional differentiation of (\ref{eq:exactflow2}).

Specific two- or four-point functions are then obtained from (\ref{eq:exactG2dia}) and (\ref{eq:exactG4dia}) by attaching the respective external dashed (${\tilde \phi}$) or solid ($\phi$) lines. For instance, starting from (\ref{eq:exactflowtptp}) the corresponding flow equation reads according to (\ref{eq:exactG2dia}) in Fourier space:\footnote{We employ \begin{eqnarray}(2\pi)^{d+1} \delta^{(d+1)}\!\!\left(\sum_{i=1}^n p_i\right)\! \Gamma^{(n)}_k(p_1,p_2,\ldots,p_{n-1})\! = \!\! \int \! \prod_{i=1}^n {\rm d}^{d+1} x_i e^{i (p^0_i x^0_i - {\bf p}_i {\bf x}_i)} \Gamma^{(n)}_k(x_1,x_2,\ldots,x_n).\nonumber\end{eqnarray}}
\begin{eqnarray}
\dot{\Gamma}_{k,ab}^{{\tilde \phi}{\tilde \phi}}(p) 
= -\frac{i}{2} \int_q &\!\Bigg\{\!& \Gamma_{k,abde}^{{\tilde \phi}{\tilde \phi}{\tilde \phi}\phi}(p,-p,q) G^{\mathrm R}_{k,ef}(q) \dot{R}^{\mathrm R}_{k,fc}(q) G^{\mathrm R}_{k,cd}(q)
\nonumber\\
&\!+\!& \Gamma_{k,abde}^{{\tilde \phi}{\tilde \phi}\phi\phi}(p,-p,q) G^{\mathrm R}_{k,ef}(q) \dot{R}^{\mathrm R}_{k,fc}(q) iF_{k,cd}(q)
\nonumber\\
&\!+\!& \Gamma_{k,abde}^{{\tilde \phi}{\tilde \phi}\phi {\tilde \phi}}(p,-p,q) G^{\mathrm A}_{k,ef}(q) \dot{R}^{\mathrm A}_{k,fc}(q) G^{\mathrm A}_{k,cd}(q)
\nonumber\\
&\!+\!& \Gamma_{k,abde}^{{\tilde \phi}{\tilde \phi}\phi\phi}(p,-p,q) iF_{k,ef}(q) \dot{R}^{\mathrm A}_{k,fc}(q) G^{\mathrm A}_{k,cd}(q) \Bigg\}. \quad
\label{eq:G2Fourier}
\end{eqnarray}
where $\int_q \equiv \int {\rm d}^{d+1} q/(2 \pi)^{d+1}$. 

The flow equations can be simplified considerably using specific representations of the cutoff functions $R_k^{\mathrm R,A}$, such as a sharp cutoff discussed in Appendix \ref{sec:sharpcutoff}.
There we show how the latter can be employed to integrate the flow analytically. This will be, in particular, important for more sophisticated approximations than the one we employ in Sec.~\ref{sec:fixedpoints}, and we give the respective formulae in the Appendix for completeness. For our case, it is advantageous to proceed without restricting the calculations to a specifc cutoff and to note that (\ref{eq:exactflow2}) can also be written, using (\ref{eq:derivR}), as
\begin{eqnarray}
\dot{\Gamma}_k[\phi,{\tilde \phi}] &\! = \!& \int_{xy} \left\{ \dot{R}^{\mathrm R}_{k,ab}(x,y)
\frac{\delta}{\delta R^{\mathrm R}_{k,ab}(x,y)} + \dot{R}^{\mathrm A}_{k,ab}(x,y)
\frac{\delta}{\delta R^{\mathrm A}_{k,ab}(x,y)} \right\} \Gamma_k[\phi,{\tilde \phi}] \, .
\nonumber\\
&\! = \!& {\rm Tr}\left\{\dot{R}_k^{\mathrm R}\frac{\delta}{\delta R^{\mathrm A}_k}
+ \dot{R}_k^{\mathrm A}\frac{\delta}{\delta R^{\mathrm R}_k} 
\right\} \Gamma_k[\phi,{\tilde \phi}] \, \equiv \, \tilde{\partial}_k\, \Gamma_k[\phi,{\tilde \phi}] .
\end{eqnarray}
For instance, the r.h.s.\ of the flow equation for the two-point function (\ref{eq:exactG2dia}) can be rewritten using relations such as  
\begin{equation}
\left( \Gamma_k^{\tilde{\phi}\phi} + R_k^{\mathrm R} \right)^{-1} \dot{R}^{\mathrm R}_k \left( \Gamma_k^{\tilde{\phi}\phi} + R_k^{\mathrm R} \right)^{-1} 
= - \tilde{\partial}_k \left( \Gamma_k^{\tilde{\phi}\phi} + R_k^{\mathrm R} \right)^{-1}. 
\label{eq:intGR}
\end{equation}
Similar relations are typically employed in a reversed sense, i.e.\ to derive renormalization group equations for $n$-point functions using diagrammatic one-loop techniques~\cite{Reviews,Pawlowski:2005xe,nonpert-reac-diff}. In order to simplify the notation, we use $G_k^{\mathrm R,A} = G_k^{\mathrm R,A}[\phi = 0,{\tilde \phi} = 0]$ and $iF_k = iF_k[\phi = 0,{\tilde \phi} = 0]$, such that the retarded/advanced and statistical propagators denote the expressions (\ref{eq:dia}). For instance, the equation for advanced functions corresponding to (\ref{eq:intGR}) reads then
\begin{equation}
G^{\mathrm A}_k \dot{R}^{\mathrm A}_k G^{\mathrm A}_k \,=\, \tilde{\partial}_k\, G^{\mathrm A}_k \, ,
\label{eq:intGA}
\end{equation}
and we have
\begin{equation}
G^{\mathrm R}_k \dot{R}^{\mathrm R}_k iF_k + iF_k \dot{R}^{\mathrm A}_k G^{\mathrm A}_k \,=\, \tilde{\partial}_k \, iF_k \, .
\label{eq:intF}
\end{equation}
The flow equation for the four-point function (\ref{eq:exactG4dia}) involves
\begin{equation}
G^{\mathrm A}_k \dot{R}^{\mathrm A}_k G^{\mathrm A}_k \Gamma^{(4)}_k G^{\mathrm R}_k \Gamma^{(4)}_k + G^{\mathrm A}_k \Gamma^{(4)}_k G^{\mathrm R}_k \dot{R}^{\mathrm R}_k G^{\mathrm R}_k \Gamma^{(4)}_k 
\, =\, \tilde{\partial}_k \left(G^{\mathrm A}_k \Gamma^{(4)}_k G^{\mathrm R}_k \Gamma^{(4)}_k \right),
\label{eq:intfourpoint1}
\end{equation}
and
\begin{eqnarray}
&&\!\!\!\!\!\! G^{\mathrm R}_k \dot{R}^{\mathrm R}_k G^{\mathrm R}_k \Gamma^{(4)}_k iF_k \Gamma^{(4)}_k + G^{\mathrm R}_k \Gamma^{(4)}_k G^{\mathrm R}_k \dot{R}^{\mathrm R}_k iF_k \Gamma^{(4)}_k + G^{\mathrm R}_k \Gamma^{(4)}_k iF_k \dot{R}^{\mathrm A}_k G^{\mathrm A}_k 
\nonumber\\
&&\!\!\!\!\!\! = \tilde{\partial}_k \left( G^{\mathrm R}_k \Gamma^{(4)}_k iF_k \Gamma^{(4)}_k \right) ,
\label{eq:intfourpoint2}
\end{eqnarray}
as well as
\begin{eqnarray}
&&\!\!\!\!\!\! \left( G^{\mathrm R}_k \dot{R}^{\mathrm R}_k iF_k + iF_k \dot{R}^{\mathrm A}_k G^{\mathrm A}_k \right) \Gamma^{(4)}_k iF_k \Gamma^{(4)}_k + iF_k \Gamma^{(4)}_k \left( G^{\mathrm R}_k \dot{R}^{\mathrm R}_k iF_k + iF_k \dot{R}^{\mathrm A}_k G^{\mathrm A}_k \right)
\nonumber\\
&&\!\!\!\!\!\! = \tilde{\partial}_k \left( iF_k \Gamma^{(4)}_k iF_k \Gamma^{(4)}_k \right) .
\label{eq:intfourpoint3}
\end{eqnarray}
  
Applying (\ref{eq:intGR})-(\ref{eq:intF}) the flow equation for the two-point function (\ref{eq:exactG2dia}) is given by
\begin{equation}
\begin{picture}(249,34) (6,-56)
    \SetWidth{0.5}
    \SetColor{Black}
    \Text(140,-42)[lb]{$\displaystyle +$}
    \Text(190,-42)[lb]{$\displaystyle +$}
    \Text(238,-47)[lb]{$\Big\} \, .$}
    \Text(10,-47)[lb]{$\displaystyle {\dot \Gamma}_{k,ab}^{(2)} 
    \,\, =\,\, -\, \frac{i}{2} \,\, {\tilde \partial}_k \,\, \Big\{$}
  \Text(110,-56)[lb]{$a$}
  \Text(127,-56)[lb]{$b$}
  \Text(160,-56)[lb]{$a$}
  \Text(177,-56)[lb]{$b$}
  \Text(210,-56)[lb]{$a$}
  \Text(227,-56)[lb]{$b$}
    \SetWidth{1.0}
    \DashCArc(120,-34)(10,90,-90){2}
    \CArc(120,-34)(10,-90,90)
    \SetWidth{0.5}
    \Vertex(120,-44){3.4}
    \SetWidth{1.0}
    \Line(110,-45)(130,-45)
    \CArc(170,-34)(10,90,270)
    \DashCArc(170,-34)(10,-90,90){2}
   \SetWidth{0.5}
    \Vertex(170,-44){3.4}
    \SetWidth{1.0}
    \Line(160,-45)(180,-45)
   \SetWidth{1.0}
    \CArc(220,-34)(10,90,-90)
    \CArc(220,-34)(10,-90,90)
    \SetWidth{0.5}
    \Vertex(220,-44){3.4}
    \SetWidth{1.0}
    \Line(210,-45)(230,-45)
  \end{picture}
\label{eq:exactoneloopG2}
\end{equation} 
Like before, equations for specific two-point functions can be directly deduced from (\ref{eq:exactoneloopG2}) by attaching the respective external dashed (${\tilde \phi}$) or solid ($\phi$) lines. For instance, in Fourier space the two-point flow equation (\ref{eq:G2Fourier}) now reads
\begin{eqnarray}
\dot{\Gamma}_{k,ab}^{{\tilde \phi}{\tilde \phi}}(p) 
= -\frac{i}{2}\,\, \tilde{\partial}_k \int_q &\!\Bigg\{\!& \Gamma_{k,abcd}^{{\tilde \phi}{\tilde \phi}{\tilde \phi}\phi}(p,-p,q) G^{\mathrm R}_{k,dc}(q) 
\nonumber\\
&\!+\!& \Gamma_{k,abcd}^{{\tilde \phi}{\tilde \phi}\phi {\tilde \phi}}(p,-p,q) G^{\mathrm A}_{k,dc}(q) 
\nonumber\\
&\!+\!& \Gamma_{k,abcd}^{{\tilde \phi}{\tilde \phi}\phi\phi}(p,-p,q) iF_{k,dc}(q)
\Bigg\}. \quad
\label{eq:G2FourierInt}
\end{eqnarray}
The four-point vertex flow equation (\ref{eq:exactG4dia}) with (\ref{eq:intfourpoint1})-(\ref{eq:intfourpoint3}) yields
\begin{eqnarray}
&&\!\!\!\!\!\!\!\!\!\!
\begin{picture}(249,49) (4,-68)
    \SetWidth{0.5}
    \SetColor{Black}
    \Text(135,-37)[lb]{$\displaystyle +$}
    \Text(215,-37)[lb]{$\displaystyle +$}
    \Text(-35,-45)[lb]{$\displaystyle \dot{\Gamma}_{k,abcd}^{(4)} 
    \,\, =\,\,  -\frac{i}{16} \,\, \tilde{\partial}_k\,\, \Big\{$}
    \Text(75,-28)[lb]{$a$}
    \Text(75,-47)[lb]{$b$}
    \Text(120,-28)[lb]{$c$}
    \Text(120,-47)[lb]{$d$}
    \Text(155,-28)[lb]{$a$}
    \Text(155,-47)[lb]{$b$}
    \Text(200,-28)[lb]{$c$}
    \Text(200,-47)[lb]{$d$}
    \Text(235,-28)[lb]{$a$}
    \Text(235,-47)[lb]{$b$}
    \Text(280,-28)[lb]{$c$}
    \Text(280,-47)[lb]{$d$}
    \SetWidth{1.0}
    \CArc(100,-34)(10,-90,90)
    \DashCArc(100,-34)(10,90,270){2}
    \SetWidth{0.5}
    \Vertex(90,-34){3.4}
    \Vertex(110,-34){3.4}
    \SetWidth{1.0}
    \Line(90,-34)(82,-25)
    \Line(90,-34)(82,-43)
    \Line(110,-34)(118,-25)
    \Line(110,-34)(118,-43)
    \DashCArc(180,-34)(10,-90,0){2}
    \CArc(180,-34)(10,180,-90)
    \DashCArc(180,-34)(10,90,180){2}
    \CArc(180,-34)(10,0,90)
    \SetWidth{0.5}
    \Vertex(170,-34){3.4}
    \Vertex(190,-34){3.4}
    \SetWidth{1.0}
    \Line(170,-34)(162,-25)
    \Line(170,-34)(162,-43)
    \Line(190,-34)(198,-25)
    \Line(190,-34)(198,-43)
    \CArc(260,-34)(10,180,90)
    \DashCArc(260,-34)(10,90,180){2}
    \SetWidth{0.5}
    \Vertex(250,-34){3.4}
    \Vertex(270,-34){3.4}
    \SetWidth{1.0}
    \Line(250,-34)(242,-25)
    \Line(250,-34)(242,-43)
    \Line(270,-34)(278,-25)
    \Line(270,-34)(278,-43)
  \end{picture}
\nonumber\\
&&\!\!\!\!\!\!\!\!\!\!  
\begin{picture}(249,30) (4,-65)
    \SetWidth{0.5}
    \SetColor{Black}
    \Text(135,-37)[lb]{$\displaystyle +$}
    \Text(215,-37)[lb]{$\displaystyle +$}
    \Text(62,-37)[lb]{$\displaystyle +$}
    \Text(75,-28)[lb]{$a$}
    \Text(75,-47)[lb]{$b$}
    \Text(120,-28)[lb]{$c$}
    \Text(120,-47)[lb]{$d$}
    \Text(155,-28)[lb]{$a$}
    \Text(155,-47)[lb]{$b$}
    \Text(200,-28)[lb]{$c$}
    \Text(200,-47)[lb]{$d$}
    \Text(235,-28)[lb]{$a$}
    \Text(235,-47)[lb]{$b$}
    \Text(280,-28)[lb]{$c$}
    \Text(280,-47)[lb]{$d$}
    \SetWidth{1.0}
    \DashCArc(100,-34)(10,0,90){2}
    \CArc(100,-34)(10,-90,0)
    \DashCArc(100,-34)(10,180,-90){2}
    \CArc(100,-34)(10,90,180)
    \SetWidth{0.5}
    \Vertex(90,-34){3.4}
    \Vertex(110,-34){3.4}
    \SetWidth{1.0}
    \Line(90,-34)(82,-25)
    \Line(90,-34)(82,-43)
    \Line(110,-34)(118,-25)
    \Line(110,-34)(118,-43)
    \DashCArc(180,-34)(10,-90,90){2}
    \CArc(180,-34)(10,90,-90)
    \SetWidth{0.5}
    \Vertex(170,-34){3.4}
    \Vertex(190,-34){3.4}
    \SetWidth{1.0}
    \Line(170,-34)(162,-25)
    \Line(170,-34)(162,-43)
    \Line(190,-34)(198,-25)
    \Line(190,-34)(198,-43)
    \CArc(260,-34)(10,90,0)
    \DashCArc(260,-34)(10,0,90){2}
    \SetWidth{0.5}
    \Vertex(250,-34){3.4}
    \Vertex(270,-34){3.4}
    \SetWidth{1.0}
    \Line(250,-34)(242,-25)
    \Line(250,-34)(242,-43)
    \Line(270,-34)(278,-25)
    \Line(270,-34)(278,-43)
  \end{picture}
\nonumber\\
&&\!\!\!\!\!\!\!\!\!\!
\begin{picture}(249,33) (4,-65)
    \SetWidth{0.5}
    \SetColor{Black}
    \Text(135,-37)[lb]{$\displaystyle +$}
    \Text(215,-37)[lb]{$\displaystyle +$}
    \Text(62,-37)[lb]{$\displaystyle +$}
    \Text(75,-28)[lb]{$a$}
    \Text(75,-47)[lb]{$b$}
    \Text(120,-28)[lb]{$c$}
    \Text(120,-47)[lb]{$d$}
    \Text(155,-28)[lb]{$a$}
    \Text(155,-47)[lb]{$b$}
    \Text(200,-28)[lb]{$c$}
    \Text(200,-47)[lb]{$d$}
    \Text(235,-28)[lb]{$a$}
    \Text(235,-47)[lb]{$b$}
    \Text(280,-28)[lb]{$c$}
    \Text(280,-47)[lb]{$d$}
    \SetWidth{1.0}
    \DashCArc(100,-34)(10,0,90){2}
    \CArc(100,-34)(10,-90,180)
    \DashCArc(100,-34)(10,180,-90){2}
    \SetWidth{0.5}
    \Vertex(90,-34){3.4}
    \Vertex(110,-34){3.4}
    \SetWidth{1.0}
    \Line(90,-34)(82,-25)
    \Line(90,-34)(82,-43)
    \Line(110,-34)(118,-25)
    \Line(110,-34)(118,-43)
    \DashCArc(180,-34)(10,-90,0){2}
    \CArc(180,-34)(10,0,-90)
    \SetWidth{0.5}
    \Vertex(170,-34){3.4}
    \Vertex(190,-34){3.4}
    \SetWidth{1.0}
    \Line(170,-34)(162,-25)
    \Line(170,-34)(162,-43)
    \Line(190,-34)(198,-25)
    \Line(190,-34)(198,-43)
    \CArc(260,-34)(10,90,0)
    \CArc(260,-34)(10,0,90)
    \SetWidth{0.5}
    \Vertex(250,-34){3.4}
    \Vertex(270,-34){3.4}
    \SetWidth{1.0}
    \Line(250,-34)(242,-25)
    \Line(250,-34)(242,-43)
    \Line(270,-34)(278,-25)
    \Line(270,-34)(278,-43)
  \end{picture}
\nonumber\\
&&\!\!\!\!\!\!\!\!\!\!
\begin{picture}(249,20) (-6,-56)
  \SetWidth{0.5}
  \SetColor{Black}
  \Text(52,-44)[lb]{$\displaystyle + \quad P(a,b,c,d) \, \Big\}$}
\end{picture}
\nonumber\\
&&\!\!\!\!\!\!\!\!\!\!
\begin{picture}(249,28) (9,-62)
    \SetWidth{0.5}
    \SetColor{Black}
    \Text(115,-44)[lb]{$\displaystyle +$}
    \Text(175,-44)[lb]{$\displaystyle +$}
    \Text(230,-50)[lb]{$\Big\} \, .$}
    \Text(25,-50)[lb]{$\displaystyle - \frac{i}{2}\,\, \tilde{\partial}_k\,\, \Big\{$}
\Text(73,-63)[lb]{$a$}
\Text(83,-63)[lb]{$b$}
\Text(93,-63)[lb]{$c$}
\Text(102,-63)[lb]{$d$}
\Text(133,-63)[lb]{$a$}
\Text(143,-63)[lb]{$b$}
\Text(153,-63)[lb]{$c$}
\Text(162,-63)[lb]{$d$}
\Text(193,-63)[lb]{$a$}
\Text(203,-63)[lb]{$b$}
\Text(213,-63)[lb]{$c$}
\Text(222,-63)[lb]{$d$}
\SetWidth{1.0}
    \CArc(90,-34)(10,-90,90)
    \DashCArc(90,-34)(10,90,270){2}
    \SetWidth{0.5}
    \Vertex(90,-44){3.4}
    \SetWidth{1.0}
    \Line(77,-55)(90,-44)
    \Line(85,-55)(90,-44)
    \Line(95,-55)(90,-44)
    \Line(103,-55)(90,-44)
    \CArc(150,-34)(10,90,270)
    \DashCArc(150,-34)(10,-90,90){2}
   \SetWidth{0.5}
    \Vertex(150,-44){3.4}
    \SetWidth{1.0}
    \Line(137,-55)(150,-44)
    \Line(145,-55)(150,-44)
    \Line(155,-55)(150,-44)
    \Line(163,-55)(150,-44)
    \CArc(210,-34)(10,90,270)
    \CArc(210,-34)(10,-90,90)
    \SetWidth{0.5}
    \Vertex(210,-44){3.4}
    \SetWidth{1.0}
    \Line(197,-55)(210,-44)
    \Line(205,-55)(210,-44)
    \Line(215,-55)(210,-44)
    \Line(223,-55)(210,-44)
  \end{picture}
\label{eq:exactoneloopG4}
\end{eqnarray} 
Again, we display all combinations of retarded, advanced and statistical propagators separately for illustration, though several obvious graphs are identical or vanish and (\ref{eq:exactoneloopG4}) can be given a more compact form. This is discussed in more detail in Sec.~\ref{sec:fixedpoints}, where an approximation in terms of a resummed $1/N$ expansion to NLO is employed. 

The above equations for $\dot{\Gamma}_k^{(2)}$ and $\dot{\Gamma}_k^{(4)}$ have to be integrated from the ultraviolet scale $\Lambda$, with $\Gamma_{k=\Lambda} = S$, to any given infrared scale $k$, and $\Gamma_{k=0} = \Gamma$ as discussed in Sec.~\ref{sec:exactflow}. For the quantum theory with action described by (\ref{eq:freeS}) and (\ref{eq:SqPhiXi}) the only nonvanishing two- and four-point functions at the ultraviolet scale are
\begin{equation}
\Gamma^{{\tilde \phi}\phi}_{\Lambda,ab}(x,y) = \Gamma^{\phi{\tilde \phi}}_{\Lambda,ab}(x,y) =
i D^{-1}_{ab}(x,y),
\end{equation}
with the classical inverse propagator (\ref{eq:classprop}), and interaction vertices
\begin{eqnarray}
\Gamma^{{\tilde \phi}\phi\phi\phi}_{\Lambda,abcd}(x,y,z,w) &\!=\!& -\frac{\lambda}{3N} \left( 
\delta_{ab} \delta_{cd} + \delta_{ac} \delta_{bd} + \delta_{bc} \delta_{ad} \right)
\nonumber\\
&\!\times\!&  \delta^{(d+1)}(x-y)\delta^{(d+1)}(x-z)\delta^{(d+1)}(x-w),
\label{eq:classicalvertex}\\
\Gamma^{{\tilde \phi}{\tilde \phi}{\tilde \phi}\phi}_{\Lambda,abcd}(x,y,z,w) &\!=\!& \frac{1}{4}\, \Gamma^{{\tilde \phi}\phi\phi\phi}_{\Lambda,abcd}(x,y,z,w) .  
\label{eq:quantumvertex}
\end{eqnarray}
Following the discussion of Sec.~\ref{sec:genfunc}, we note that for the corresponding classical-statistical field theory the "quantum vertex" $\Gamma^{{\tilde \phi}{\tilde \phi}{\tilde \phi}\phi}_{\Lambda}$ vanishes according to (\ref{eq:ClSqPhiXi}). In contrast, in the quantum theory it is only suppressed by a factor of four compared to $\Gamma^{{\tilde \phi}\phi\phi\phi}_{\Lambda}$. The latter is present in both quantum-statistical and classical-statistical field theories~\cite{Berges:2007ym,MSR}.

\section{Nonthermal fixed points}
\label{sec:fixedpoints}

\subsection{Stationarity condition}
\label{sec:scaling}

Infrared fixed points correspond to low-momentum scaling solutions for $n$-point correlation functions. 
As mentioned in Sec.~\ref{sec:introfp}, the absence of a fluctuation-dissipation relation out of equilibrium can lead to new nonthermal scaling solutions. This was pointed out recently in Ref.~\cite{Berges:2008wm} for the theory considered here, based on a stationarity condition for time translation invariant solutions of nonequilibrium time evolution equations~\cite{Berges:2004yj}.
An equivalent (scale-dependent) stationarity condition can be obtained from the functional renormalization group, where it appears as a nontrivial identity relating the various second functional derivatives of $\Gamma_k[\phi,{\tilde \phi}]$. For this we consider 
\begin{eqnarray}
i \Gamma^{\tilde{\phi}\tilde{\phi}}_{k,ac}\left\{ G^{\mathrm R}_{k,ca} - G^{\mathrm A}_{k,ca} \right\} 
&=& -i G^{\mathrm R}_{k,ac} \Gamma^{\tilde{\phi}\tilde{\phi}}_{k,cd} G^{\mathrm A}_{k,de}
\left\{ \left( G^{\mathrm R}_k \right)^{-1}_{ea} - \left( G^{\mathrm A}_k \right)^{-1}_{ea} \right\} 
\nonumber\\
&=& F_{k,ac} \left\{ \Gamma^{\phi\tilde{\phi}}_{k,ca} - \Gamma^{\tilde{\phi}\phi}_{k,ca} \right\}\, , 
\label{eq:stationarity}
\end{eqnarray}  
where we suppress space-time dependencies in the notation.
The first line of (\ref{eq:stationarity}) can be verified by simple matrix multiplication, and the second line follows from (\ref{eq:Rchoice}) and (\ref{eq:dia}). Following conventional notations, 
$i \Gamma^{\tilde{\phi}\tilde{\phi}}_k$ represents the statistical part and $\Gamma^{\phi\tilde{\phi}}_k - \Gamma^{\tilde{\phi}\phi}_k$ the spectral part of the self-energy denoted as~\cite{Berges:2007ym} 
\begin{equation}
\Sigma^F_{k,ab} \equiv i \Gamma^{\tilde{\phi}\tilde{\phi}}_{k,ab} \quad,\quad
\Sigma^{\rho}_{k,ab} \equiv \Gamma^{\phi\tilde{\phi}}_{k,ab} - \Gamma^{\tilde{\phi}\phi}_{k,ab} \, .
\label{eq:SFSR}
\end{equation}
The latter is related to the retarded/advanced self-energy by 
$\Sigma^{\rho}_{k,ab} = \Sigma^{\mathrm R}_{k,ab} - \Sigma^{\mathrm A}_{k,ab}$, similar to the spectral function being related to the retarded/advanced propagator by $\rho_{k,ab} = G^{\mathrm R}_{k,ab} - G^{\mathrm A}_{k,ab}$ according to (\ref{eq:RGRGA}). 
Using this notation (\ref{eq:stationarity}) reads in Fourier space 
\begin{equation}
\Sigma^F_{k,ac}(p) \rho_{k,ca}(p) = F_{k,ac}(p) \Sigma^{\rho}_{k,ca}(p) \, . 
\label{eq:2PIstat}
\end{equation}
This equation is well-known in nonequilibrium physics. In the language of Boltzmann dynamics it states that "gain terms" equal "loss terms" for which stationarity is achieved~\cite{Berges:2004yj}.

The flow equations for the self-energies are determined with (\ref{eq:SFSR}) and (\ref{eq:exactoneloopG2}): 
\begin{eqnarray}
\dot{\Sigma}^F_{k,ab}(p)
&\!=\!& \frac{1}{2}\, \tilde{\partial}_k \int_q \Bigg\{ \Gamma_{k,abcd}^{{\tilde \phi}{\tilde \phi}{\tilde \phi}\phi}(p,-p,q) G^{\mathrm R}_{k,dc}(q) 
+ \Gamma_{k,abcd}^{{\tilde \phi}{\tilde \phi}\phi {\tilde \phi}}(p,-p,q) G^{\mathrm A}_{k,dc}(q) 
\nonumber\\
&& +\, \Gamma_{k,abcd}^{{\tilde \phi}{\tilde \phi}\phi\phi}(p,-p,q) iF_{k,dc}(q)
\Bigg\},
\label{eq:SigF}
\\
\dot{\Sigma}^{\rho}_{k,ab}(p)
&\!=\!& - \frac{i}{2}\, \tilde{\partial}_k \int_q \Bigg\{ \left[ \Gamma_{k,abcd}^{\phi{\tilde \phi}{\tilde \phi}\phi}(p,-p,q) - \Gamma_{k,abcd}^{{\tilde \phi}\phi{\tilde \phi}\phi}(p,-p,q) \right] G^{\mathrm R}_{k,dc}(q) 
\nonumber\\
&& + \left[ \Gamma_{k,abcd}^{\phi {\tilde \phi}\phi {\tilde \phi}}(p,-p,q) - \Gamma_{k,abcd}^{{\tilde \phi} \phi \phi {\tilde \phi}}(p,-p,q) \right] G^{\mathrm A}_{k,dc}(q) 
\nonumber\\
&& + \left[ \Gamma_{k,abcd}^{\phi {\tilde \phi}\phi\phi}(p,-p,q) - \Gamma_{k,abcd}^{{\tilde \phi} \phi \phi \phi}(p,-p,q) \right] iF_{k,dc}(q)
\Bigg\}. 
\label{eq:SigR}
\end{eqnarray}
For given four-point functions the stationarity equation (\ref{eq:2PIstat}) with (\ref{eq:SigF}) and (\ref{eq:SigR}) depends on $F_{k,ab}(p)$ and $G^{\mathrm R}_{k,ab}(p) = G^{\mathrm A}_{k,ba}(-p)$ or, equivalently, $\rho_{k,ab}(p)$. We have described in Sec.~\ref{sec:introfp} that the scaling behavior of these two-point functions at an infrared fixed point may be expressed using the occupation number exponent $\kappa$, dynamical scaling exponent z and anomalous dimension $\eta$. Using the $O(N)$ symmetry for $\phi = {\tilde \phi} = 0$ their scaling forms read\footnote{For $k \equiv 0$ the scaling properties are equivalently described by (\ref{eq:scaling}).} 
\begin{eqnarray}
F_{k,ab}(p^0,{\bf p}) &\!=\!& \frac{1}{k^{2+\kappa}}\,\, f\!\left( \frac{p^0}{k^z},\frac{|{\bf p}|}{k}\right) \delta_{ab} \, , 
\nonumber\\
G^{\mathrm R,A}_{k,ab}(p^0,{\bf p}) &\!=\!& \frac{1}{k^{2-\eta}}\,\, g^{\mathrm R,A}\!\left( \frac{p^0}{k^z},\frac{|{\bf p}|}{k}\right) \delta_{ab}\, ,
\label{eq:kscalingform}
\end{eqnarray}
where the scaling functions $f$ and $g^{\mathrm R,A}$ only depend on the scaling ratios of the frequency $p^0$, the absolute value of spatial momentum $|{\bf p}|$ and the scale $k$. In thermal equilibrium the fluctuation-dissipation relation implies the relation (\ref{eq:vacuumconst}) or (\ref{eq:constraint}) for $\kappa$, $z$ and $\eta$. As an application of the nonequilibrium renormalization group method, we will determine in the following corresponding relations between these exponents at nonthermal fixed points using a nonperturbative large-$N$ expansion to NLO.

\subsection{Resummed $1/N$ expansion to next-to-leading order}
\label{sec:2PIlargeN}

The search for infrared scaling solutions of the equation for two-point functions (\ref{eq:exactG2dia}) or (\ref{eq:exactoneloopG2}) requires a simultaneous solution for four-point functions 
(\ref{eq:exactG4dia}) or (\ref{eq:exactoneloopG4}), which itself depend on six-point functions etc. In order to truncate this infinite hierarchy of equations we employ a resummed $1/N$ expansion to NLO. A similar calculation for time evolution equations can be found in Ref.~\cite{Gasenzer:2007za}. We write
\begin{equation}
\Gamma^{{\tilde \phi} \phi}_{k,ab}(x,y) \,=\, \Gamma^{{\tilde \phi} \phi}_k(x,y) \delta_{ab} 
\label{eq:diag2}
\end{equation} 
and
\begin{eqnarray}
\Gamma^{{\tilde \phi} \phi\phi\phi}_{k,abcd}(x,y,z,w) &\!=\!&
\Gamma^{{\tilde \phi} \phi,\phi\phi}_{k}(x,z)\, \delta^{(d+1)}(x-y)\, \delta^{(d+1)}(z-w) \,\delta_{ab}\delta_{cd}
\nonumber\\
&\! + \!& \Gamma^{{\tilde \phi} \phi,\phi\phi}_{k}(z,y)\, \delta^{(d+1)}(z-x)\, \delta^{(d+1)}(y-w)\, \delta_{ac}\delta_{bd}
\nonumber\\
&\! + \!& \Gamma^{\phi\phi,{\tilde \phi} \phi}_{k}(y,x) \delta^{(d+1)}(y-z)\, \delta^{(d+1)}(x-w)\, \delta_{bc}\delta_{ad}
\nonumber\\
&\! + \!& \Gamma^{{\tilde \phi} \phi\phi\phi}_{\Lambda}\, \delta^{(d+1)}(x-y)\, \delta^{(d+1)}(x-z)\, \delta^{(d+1)}(x-w)
\nonumber\\
&\!  \!&  \times \left(\delta_{ab}\delta_{cd}+\delta_{ac}\delta_{bd}+\delta_{bc}\delta_{ad} \right) \, .
\label{eq:diag4}
\end{eqnarray}
The other two- and four-derivatives are written accordingly. For a compact diagrammatic notation we will draw propagators along the lines of Sec.~\ref{sec:dia} and four-vertex functions as
\begin{equation}
  \begin{picture}(-28,43) (198,-241)
    \SetWidth{0.5}
    \SetColor{Black}
 \Text(32,-232)[lb]{$\Gamma_k^{\tilde{\phi}\phi , \phi \phi}(x,y)\,=\,$}
 \Text(190,-232)[lb]{$\Gamma_k^{\tilde{\phi}\tilde{\phi} , \tilde{\phi} \phi}(x,y) \,=\, $}
 \Text(117,-228)[lb]{$x$}
 \Text(158,-229)[lb]{$y$}
 \Text(277,-228)[lb]{$x$}
 \Text(318,-229)[lb]{$y$}
 \Text(164,-232)[lb]{\quad,}
   \SetWidth{1}
    \GOval(140,-224)(3,10)(0){0}
  \DashLine(129,-224)(121,-214){2}
  \Line(129,-224)(121,-234)
  \Line(151,-224)(159,-214)
  \Line(151,-224)(159,-234)
\SetWidth{1}
    \GOval(300,-224)(3,10)(0){0}
  \DashLine(289,-224)(281,-214){2}
  \DashLine(289,-224)(281,-234){2}
  \DashLine(311,-224)(319,-214){2}
  \Line(311,-224)(319,-234)
  \end{picture}
\end{equation}
etc. Plugging (\ref{eq:diag2}) and (\ref{eq:diag4}) into the flow equations for the self-energies (\ref{eq:SigF}) and (\ref{eq:SigR}) the contributions can be classified according to their scaling with $N$. The leading contributions $\sim N$ turn out to vanish identically. For (\ref{eq:SigF}) this is a consequence of the fact that with the vertices (\ref{eq:classicalvertex}) and (\ref{eq:quantumvertex}) any local ("tadpole-type") contribution to $\Sigma^F$ vanishes because of $\int_x G^{\mathrm R,A}(x,x) = 0$ and $\int_{x,y} G^{\mathrm R}(x,y)G^{\mathrm A}(x,y) = 0$. For (\ref{eq:SigR}) the same type of contributions cancel because of simple subtraction. Diagrammatically, one is left with the equations
\begin{eqnarray}
&& \begin{picture}(3,26) (220,-232)
    \SetWidth{0.5}
    \SetColor{Black}
  \Text(26,-234)[lb]{$\displaystyle \Sigma_k^{\rho \,} \,=\, -i \, \Big\{ $}
  \Text(146,-227)[lb]{$\displaystyle +$}
  \Text(226,-227)[lb]{$\displaystyle -$}
  \Text(306,-227)[lb]{$\displaystyle -$}
  \Text(380,-232)[lb]{$\displaystyle \Big\} \,, $}
    \SetWidth{1}
    \GOval(110,-224)(3,10)(0){0}
  \Line(88,-224)(98,-224)
  \DashLine(132,-224)(122,-224){2}
  \CArc(110,-224)(12,0,90)
  \DashCArc(110,-224)(12,90,180){2}
      \SetWidth{1}
    \GOval(190,-224)(3,10)(0){0}
  \Line(168,-224)(178,-224)
  \DashLine(212,-224)(202,-224){2}
  \CArc(190,-224)(12,0,180)
     \SetWidth{1}
    \GOval(270,-224)(3,10)(0){0}
  \DashLine(248,-224)(258,-224){2}
  \Line(292,-224)(282,-224)
  \DashCArc(270,-224)(12,0,90){2}
  \CArc(270,-224)(12,90,180)
  \SetWidth{1}
    \GOval(350,-224)(3,10)(0){0}
  \DashLine(328,-224)(338,-224){2}
  \Line(372,-224)(362,-224)
  \CArc(350,-224)(12,0,180)
 \end{picture}
\nonumber\\
&& \begin{picture}(3,26) (220,-232)
    \SetWidth{0.5}
    \SetColor{Black}
  \Text(25,-224)[lb]{$\displaystyle \Sigma_k^{F} \,=\, $}
  \Text(146,-225)[lb]{$\displaystyle +$}
  \Text(226,-225)[lb]{$\displaystyle +$}
  \Text(306,-227)[lb]{$\displaystyle .$}
    \SetWidth{1}
    \GOval(110,-224)(3,10)(0){0}
  \DashLine(88,-224)(98,-224){2}
  \DashLine(132,-224)(122,-224){2}
  \CArc(110,-224)(12,0,90)
  \DashCArc(110,-224)(12,90,180){2}
      \SetWidth{1}
    \GOval(190,-224)(3,10)(0){0}
  \DashLine(168,-224)(178,-224){2}
  \DashLine(212,-224)(202,-224){2}
  \DashCArc(190,-224)(12,0,90){2}
  \CArc(190,-224)(12,90,180)
  \SetWidth{1}
    \GOval(270,-224)(3,10)(0){0}
  \DashLine(248,-224)(258,-224){2}
  \DashLine(292,-224)(282,-224){2}
  \CArc(270,-224)(12,0,180)
  \end{picture}
\label{eq:selfdiag}
\end{eqnarray}
We emphasize that in (\ref{eq:selfdiag}) we wrote the integrated form of the flow equations, which "dropped" the scale derivatives appearing in (\ref{eq:SigF}) and (\ref{eq:SigR}). This trivial integration is possible because they represent total derivatives to this order in the $1/N$ expansion as is explained below.    

For the following it is convenient to introduce one-loop self-energies with dressed propagators given by\footnote{We note that in Fourier space $F_k(p)$ is real while $\rho_k(p)$ is purely imaginary.}
\begin{eqnarray}
\Pi^{\mathrm R,A}_k(p) & \! = \! & \frac{\lambda}{3} \int_q F_k(p-q) G^{\mathrm R,A}_k(q)\quad ,
\quad \Pi^{\rho}_k(p) \, = \, \Pi^{\mathrm R}_k(p) - \Pi^{\mathrm A}_k(p) \, ,
\nonumber\\
\Pi^F_k(p) & \! = \! & \frac{\lambda}{6} \int_q \left[ F_k(p-q) F_k(q) - \frac{1}{4} \rho_k(p-q) \rho_k(q) \right] \, ,
\label{eq:quantumPi}
\end{eqnarray}
and a momentum-dependent "effective coupling" $\lambda_{{\rm eff},k}(p)$ with
\begin{eqnarray}
\frac{\lambda_{{\rm eff},k}(p)}{\lambda} & \! = \! & \frac{1}{\left[1+\Pi^{\mathrm R}_k(p)\right]\left[1+\Pi^{\mathrm A}_k(p)\right]} 
\nonumber\\
& = & 1 - \Pi^{\mathrm R}_k(p) - \Pi^{\mathrm A}_k(p) + \left(\Pi^{\mathrm R}_k(p)\right)^2
+ \left(\Pi^{\mathrm A}_k(p)\right)^2 + \ldots
\, .
\label{eq:effcoup}
\end{eqnarray} 
For the second line in (\ref{eq:effcoup}) we have used that powers of mixed retarded/advanced 
products $\Pi^{\mathrm R}_k(p) \Pi^{\mathrm A}_k(p)$ vanish identically.

The six-point vertex contributes to the flow of $\Gamma_k^{(4)}$ only at subleading order in the $1/N$ expansion. The leading contribution to the flow described by (\ref{eq:exactoneloopG4}) with (\ref{eq:diag4}) generates an infinite chain of "bubble" diagrams with dressed propagators when iterated about the short-distance vertices (\ref{eq:classicalvertex}) and (\ref{eq:quantumvertex}). The $k$-integral of this set of flow equations then contains all bubble-chain contributions and is trivially obtained from (\ref{eq:exactoneloopG4}) as
\begin{eqnarray}
&&
\begin{picture}(45,14) (200,-226)
\SetWidth{0.5}
    \SetColor{Black}
\Text(60,-236)[lb]{$\displaystyle = \, -\frac{i}{2}N \, \Big\{ \,$}
\Text(192,-229)[lb]{$\displaystyle + $}
\Text(282,-229)[lb]{$\displaystyle + $}
\Text(367,-236)[lb]{$\displaystyle \, \Big\} \, $}
  \SetWidth{1}
    \GOval(40,-224)(3,6)(0){0}
  \DashLine(33,-224)(25,-216){2}
  \Line(33,-224)(25,-232)
  \DashLine(47,-224)(55,-216){2}
  \Line(47,-224)(55,-232)
\SetWidth{1}
   \GOval(133,-224)(3,6)(0){0}
   \DashLine(126,-224)(118,-216){2}
   \Line(126,-224)(118,-232)
  \CArc(150,-224)(10,-90,90)
  \DashCArc(150,-224)(10,90,-90){2}
   \GOval(167,-224)(3,6)(0){0}
   \DashLine(174,-224)(182,-216){2}
   \Line(174,-224)(182,-232)
\SetWidth{1}
   \GOval(223,-224)(3,6)(0){0}
   \DashLine(216,-224)(208,-216){2}
   \Line(216,-224)(208,-232)
  \CArc(240,-224)(10,90,-90)
  \DashCArc(240,-224)(10,-90,90){2}
   \GOval(257,-224)(3,6)(0){0}
   \DashLine(264,-224)(272,-216){2}
   \Line(264,-224)(272,-232)
\SetWidth{1}
   \GOval(313,-224)(3,6)(0){0}
   \DashLine(306,-224)(298,-216){2}
   \Line(306,-224)(298,-232)
  \CArc(330,-224)(10,90,-90)
  \CArc(330,-224)(10,-90,90)
   \GOval(347,-224)(3,6)(0){0}
   \DashLine(354,-224)(362,-216){2}
   \Line(354,-224)(362,-232)
\end{picture}
\nonumber\\
&&  
\begin{picture}(45,27) (200,-226)
\SetWidth{0.5}
    \SetColor{Black}
\Text(60,-236)[lb]{$\displaystyle = \, -\frac{i}{2}N \, \Big\{ \,$}
\Text(164,-229)[lb]{$\displaystyle + $}
\Text(226,-229)[lb]{$\displaystyle + $}
\Text(285,-236)[lb]{$\displaystyle \, \Big\} \, $}
\SetWidth{1}
   \DashLine(126,-224)(118,-216){2}
   \Line(126,-224)(118,-232)
  \CArc(137,-224)(10,-90,90)
  \DashCArc(137,-224)(10,90,-90){2}
   \DashLine(148,-224)(158,-216){2}
   \Line(148,-224)(158,-232)
\SetWidth{1}
   \DashLine(188,-224)(180,-216){2}
   \Line(188,-224)(180,-232)
  \CArc(199,-224)(10,90,-90)
  \DashCArc(199,-224)(10,-90,90){2}
   \DashLine(210,-224)(218,-216){2}
   \Line(210,-224)(218,-232)
\SetWidth{1}
   \DashLine(250,-224)(242,-216){2}
   \Line(250,-224)(242,-232)
  \CArc(261,-224)(10,90,-90)
  \CArc(261,-224)(10,-90,90)
   \DashLine(272,-224)(280,-216){2}
   \Line(272,-224)(280,-232)
\end{picture}
\nonumber\\
&&
\begin{picture}(45,27) (200,-226)
\SetWidth{0.5}
    \SetColor{Black}
\Text(74,-236)[lb]{$\displaystyle + \, \frac{1}{2}(-iN)^2 \, \Big\{ \,$}
\Text(213,-229)[lb]{$\displaystyle + $}
\Text(293,-229)[lb]{$\displaystyle + $}
\SetWidth{1}
   \DashLine(156,-224)(148,-216){2}
   \Line(156,-224)(148,-232)
  \CArc(167,-224)(10,90,0)
  \DashCArc(167,-224)(10,0,90){2}
  \CArc(188,-224)(10,-90,90)
  \DashCArc(188,-224)(10,90,-90){2}
   \DashLine(199,-224)(207,-216){2}
   \Line(199,-224)(207,-232)
\SetWidth{1}
   \DashLine(236,-224)(228,-216){2}
   \Line(236,-224)(228,-232)
  \CArc(247,-224)(10,-90,90)
  \DashCArc(247,-224)(10,90,-90){2}
  \CArc(268,-224)(10,180,90)
  \DashCArc(268,-224)(10,90,180){2}
   \DashLine(279,-224)(287,-216){2}
   \Line(279,-224)(287,-232)
\SetWidth{1}
   \DashLine(316,-224)(308,-216){2}
   \Line(316,-224)(308,-232)
  \CArc(327,-224)(10,90,0)
  \DashCArc(327,-224)(10,0,90){2}
  \CArc(348,-224)(10,90,-90)
  \DashCArc(348,-224)(10,-90,90){2}
   \DashLine(359,-224)(367,-216){2}
   \Line(359,-224)(367,-232)
\end{picture}
\nonumber\\
&&  
\begin{picture}(45,27) (200,-226)
\SetWidth{0.5}
    \SetColor{Black}
\Text(133,-229)[lb]{$\displaystyle + $}
\Text(213,-229)[lb]{$\displaystyle + $}
\Text(293,-229)[lb]{$\displaystyle + $}
\Text(371,-236)[lb]{$\displaystyle \, \Big\} \, + \, ... $}
\SetWidth{1}
   \DashLine(156,-224)(148,-216){2}
   \Line(156,-224)(148,-232)
  \CArc(167,-224)(10,90,-90)
  \DashCArc(167,-224)(10,-90,90){2}
  \CArc(188,-224)(10,180,90)
  \DashCArc(188,-224)(10,90,180){2}
   \DashLine(199,-224)(207,-216){2}
   \Line(199,-224)(207,-232)
\SetWidth{1}
   \DashLine(236,-224)(228,-216){2}
   \Line(236,-224)(228,-232)
  \CArc(247,-224)(10,90,0)
  \DashCArc(247,-224)(10,0,90){2}
  \CArc(268,-224)(10,180,90)
  \CArc(268,-224)(10,90,180)
   \DashLine(279,-224)(287,-216){2}
   \Line(279,-224)(287,-232)
\SetWidth{1}
   \DashLine(316,-224)(308,-216){2}
   \Line(316,-224)(308,-232)
  \CArc(327,-224)(10,-90,90)
  \CArc(327,-224)(10,90,-90)
  \CArc(348,-224)(10,180,90)
  \DashCArc(348,-224)(10,90,180){2}
   \DashLine(359,-224)(367,-216){2}
   \Line(359,-224)(367,-232)
\end{picture}
\nonumber\\
&&  
\begin{picture}(45,32) (200,-226)
  \SetWidth{0.5}
    \SetColor{Black}
 \Text(60,-229)[lb]{$\displaystyle =\, i\, \frac{\lambda}{3N} \, \Big\{ \, 1- \Pi^{R}_k \,-\, \Pi^{A}_k \, + \, (\Pi^{R}_k)^2 \,+\, (\Pi^{A}_k)^2 \, + \, ... \, \Big\} \, \Pi^{F}_k $}
\end{picture}
\nonumber\\
&& 
\begin{picture}(45,35) (200,-228)
  \SetWidth{0.5}
    \SetColor{Black}
\Text(60,-229)[lb]{$\displaystyle =\, i\, \frac{\lambda_{{\rm eff},k}}{3N}\, \Pi^{F}_k \,$.}
\end{picture} 
\label{eq:four1}
\end{eqnarray}
The second line is derived from the first one in (\ref{eq:four1}) by simultaneously replacing the dressed vertices by their iterated expression to infinite powers in $\Gamma^{{\tilde \phi}\phi\phi\phi}_{\Lambda}$ and $\Gamma^{{\tilde \phi}{\tilde \phi}{\tilde \phi}\phi}_{\Lambda}$ defined in (\ref{eq:classicalvertex}) and (\ref{eq:quantumvertex}), respectively. The first line in the above equation depends also on $\Gamma_k^{\phi\phi, \tilde{\phi} \phi}(p)$ and $\Gamma_k^{\tilde{\phi}\tilde{\phi}, \tilde{\phi} \phi}(p)$, as well as $\Gamma_k^{\tilde{\phi} \phi,\phi\phi}(p)$ and $\Gamma_k^{\tilde{\phi} \phi, \tilde{\phi}\tilde{\phi}}(p)$. The first two are given by
\begin{eqnarray}
&&
\begin{picture}(73,12) (168,-226)
\SetWidth{0.5}
    \SetColor{Black}
\Text(78,-227)[lb]{$\displaystyle = $}
\Text(126,-229)[lb]{$\displaystyle -iN $}
  \SetWidth{1}
    \GOval(55,-224)(3,6)(0){0}
  \Line(48,-224)(40,-216)
  \Line(48,-224)(40,-232)
  \DashLine(62,-224)(70,-216){2}
  \Line(62,-224)(70,-232)
  \SetWidth{1}
  \Line(106,-224)(98,-216)
  \Line(106,-224)(98,-232)
  \DashLine(106,-224)(114,-216){2}
  \Line(106,-224)(114,-232)
  \SetWidth{1}
   \GOval(171,-224)(3,6)(0){0}
   \Line(164,-224)(156,-216)
   \Line(164,-224)(156,-232)
  \CArc(188,-224)(10,180,90)
  \DashCArc(188,-224)(10,90,180){2}
   \GOval(206,-224)(3,6)(0){0}
  \Line(213,-224)(221,-234)
  \DashLine(213,-224)(221,-214){2}
\end{picture}
\nonumber\\
&&  
\begin{picture}(73,32) (168,-236)
  \SetWidth{0.5}
    \SetColor{Black}
 \Text(78,-227)[lb]{$\displaystyle = $}
 \Text(126,-229)[lb]{$\displaystyle -iN $}
 \Text(205,-229)[lb]{$\displaystyle + \, (-iN)^2 $}
 \Text(325,-228)[lb]{$\displaystyle + \, ... $}
  \SetWidth{1}
  \Line(106,-224)(98,-216)
  \Line(106,-224)(98,-232)
  \DashLine(106,-224)(114,-216){2}
  \Line(106,-224)(114,-232)
  \SetWidth{1}
   \Line(164,-224)(156,-216)
   \Line(164,-224)(156,-232)
   \CArc(174,-224)(10,180,90)
   \DashCArc(174,-224)(10,90,180){2}
   \DashLine(184,-224)(192,-216){2}
   \Line(184,-224)(192,-232)
  \SetWidth{1}
   \Line(267,-224)(259,-214)
   \Line(267,-224)(259,-234)
   \CArc(277,-224)(10,180,90)
   \DashCArc(277,-224)(10,90,180){2}
   \CArc(298,-224)(10,180,90)
   \DashCArc(298,-224)(10,90,180){2}
   \DashLine(308,-224)(316,-216){2}
   \Line(308,-224)(316,-232)
\end{picture}
\nonumber\\
&&  
\begin{picture}(73,32) (168,-236)
  \SetWidth{0.5}
    \SetColor{Black}
 \Text(78,-229)[lb]{$\displaystyle =\, \Gamma_{\Lambda}^{\tilde{\phi}\phi\phi\phi} \, \Big\{ \, 1- \Pi^{A}_k \, + \, (\Pi^{A}_k)^2 \, + \, ... \, \Big\}$}
\end{picture}
\nonumber\\
&&
\begin{picture}(73,15) (168,-224)
  \SetWidth{0.5}
    \SetColor{Black}
 \Text(78,-229)[lb]{$\displaystyle =\, -\frac{\lambda}{3N} + \frac{\lambda_{{\rm eff},k}}{3N}\, \Pi^{A}_k $}
\end{picture} 
\label{eq:classeff}
\end{eqnarray}
and
\begin{eqnarray}
&&
\begin{picture}(73,-12) (168,-230)
\SetWidth{0.5}
    \SetColor{Black}
\Text(78,-227)[lb]{$\displaystyle = $}
\Text(126,-229)[lb]{$\displaystyle -iN $}
  \SetWidth{1}
    \GOval(55,-224)(3,6)(0){0}
  \DashLine(48,-224)(40,-216){2}
  \DashLine(48,-224)(40,-232){2}
  \DashLine(62,-224)(70,-216){2}
  \Line(62,-224)(70,-232)
  \SetWidth{1}
  \DashLine(106,-224)(98,-216){2}
  \DashLine(106,-224)(98,-232){2}
  \DashLine(106,-224)(114,-216){2}
  \Line(106,-224)(114,-232)
  \SetWidth{1}
   \GOval(171,-224)(3,6)(0){0}
   \DashLine(164,-224)(156,-216){2}
   \DashLine(164,-224)(156,-232){2}
  \CArc(188,-224)(10,180,90)
  \DashCArc(188,-224)(10,90,180){2}
   \GOval(206,-224)(3,6)(0){0}
  \Line(213,-224)(221,-234)
  \DashLine(213,-224)(221,-214){2}
\end{picture}
\nonumber\\
&&  
\begin{picture}(73,32) (168,-236)
  \SetWidth{0.5}
    \SetColor{Black}
 \Text(78,-227)[lb]{$\displaystyle = $}
 \Text(126,-229)[lb]{$\displaystyle -iN $}
 \Text(205,-229)[lb]{$\displaystyle + \, (-iN)^2 $}
 \Text(325,-228)[lb]{$\displaystyle + \, ... $}
  \SetWidth{1}
  \DashLine(106,-224)(98,-216){2}
  \DashLine(106,-224)(98,-232){2}
  \DashLine(106,-224)(114,-216){2}
  \Line(106,-224)(114,-232)
  \SetWidth{1}
   \DashLine(164,-224)(156,-216){2}
   \DashLine(164,-224)(156,-232){2}
   \CArc(174,-224)(10,180,90)
   \DashCArc(174,-224)(10,90,180){2}
   \DashLine(184,-224)(192,-216){2}
   \Line(184,-224)(192,-232)
  \SetWidth{1}
   \DashLine(267,-224)(259,-214){2}
   \DashLine(267,-224)(259,-234){2}
   \CArc(277,-224)(10,180,90)
   \DashCArc(277,-224)(10,90,180){2}
   \CArc(298,-224)(10,180,90)
   \DashCArc(298,-224)(10,90,180){2}
   \DashLine(308,-224)(316,-216){2}
   \Line(308,-224)(316,-232)
\end{picture}
\nonumber\\
&&  
\begin{picture}(73,32) (168,-236)
  \SetWidth{0.5}
    \SetColor{Black}
 \Text(78,-229)[lb]{$\displaystyle =\, \Gamma_{\Lambda}^{\tilde{\phi}\tilde{\phi}\tilde{\phi}\phi} \, \Big\{ \, 1- \Pi^{A}_k \, + \, (\Pi^{A}_k)^2 \, + \, ... \, \Big\}$}
\end{picture}
\nonumber\\
&&
\begin{picture}(73,18) (168,-226)
  \SetWidth{0.5}
    \SetColor{Black}
 \Text(78,-229)[lb]{$\displaystyle =\, -\frac{\lambda}{12N} + \frac{\lambda_{{\rm eff},k}}{12N}\, \Pi^{A}_k \, $.}
\end{picture}
\label{eq:quanteff}
\end{eqnarray}
For the last line in (\ref{eq:classeff}) and (\ref{eq:quanteff}) we use again that products of retarded and advanced self-energies at same momentum vanish. 
Similar to the short-distance vertices (\ref{eq:classicalvertex}) and (\ref{eq:quantumvertex}), the proper vertex (\ref{eq:classeff}) plays the role of a classical and (\ref{eq:quanteff}) that of a quantum vertex. Accordingly, one observes that both are related by
\begin{equation}
\Gamma_k^{\tilde{\phi}\tilde{\phi}, \tilde{\phi} \phi}(p) = \frac{1}{4}\, \Gamma_k^{\phi\phi, \tilde{\phi} \phi}(p) \, , 
\end{equation}
and the proper vertex (\ref{eq:quanteff}) is absent in a classical-statistical field theory. The same holds for the corresponding retarded functions
\begin{equation}
\Gamma_k^{\tilde{\phi} \phi, \tilde{\phi}\tilde{\phi}}(p) = -\frac{\lambda}{12N} + \frac{\lambda_{{\rm eff},k}(p)}{12N}\, \Pi^{R}_k(p) = \frac{1}{4}\, \Gamma_k^{\tilde{\phi} \phi,\phi\phi}(p) \, .
\end{equation}
With these results the self-energies (\ref{eq:selfdiag}) at NLO read
\begin{eqnarray}
\Sigma^F_k(p) &\!\! = \!\!& - \frac{1}{3N} \int_q \lambda_{{\rm eff},k}(p-q) \left\{
\Pi^F_k(p-q) F_k(q) - \frac{1}{4} \Pi^{\rho}_k(p-q) \rho_k(q)
\right\}, 
\nonumber\\
\Sigma^{\rho}_k(p) &\!\! = \!\!& - \frac{1}{3N} \int_q \lambda_{{\rm eff},k}(p-q)
\left\{
\Pi^F_k(p-q) \rho_k(q) + \Pi^{\rho}_k(p-q) F_k(q)
\right\}. 
\label{eq:NLOselfenergies}
\end{eqnarray}
We observe that this agrees for $k=0$ with the results obtained from a $1/N$ expansion of the 2PI effective action to NLO \cite{Berges:2001fi,Berges:2008wm,Gasenzer:2007za}.

\subsection{Fixed points}
\label{sec:fixedpointscalc}

We are looking for infrared scaling solutions (\ref{eq:kscalingform}) of the stationarity equation (\ref{eq:2PIstat}), using the approximation (\ref{eq:NLOselfenergies}) for the self-energies. Plugging into (\ref{eq:quantumPi}) and (\ref{eq:NLOselfenergies}) the scaling form of the propagators (\ref{eq:kscalingform}), each contribution to the self-energies can be classified according to their scaling with powers of $k$.
For $2+\kappa > 0$ and $\kappa > - \eta$ the leading infrared behavior of $\Pi^F_k(p)$ and $\Sigma^F_k(p)$ is described by  
\begin{equation}  
\Pi^F_k(p) \, \simeq \, \frac{\lambda}{6} \int_q F_k(p-q) F_k(q)
\label{eq:PiFclass} 
\end{equation}
and 
\begin{equation}
\Sigma^F_k(p) \, \simeq \, - \frac{1}{3N} \int_q \lambda_{{\rm eff},k}(p-q) 
\Pi^F_k(p-q) F_k(q) \, .
\label{eq:SigmaFclass}
\end{equation}
The spectral self-energies $\Pi^{\rho}_k(p)$ and $\Sigma^{\rho}_k(p)$ are still given by (\ref{eq:quantumPi}) and (\ref{eq:NLOselfenergies}), respectively, with $\Pi^F_k(p)$ replaced by (\ref{eq:PiFclass}). We note that the above approximation corresponds to the classical-statistical field theory limit of the underlying quantum theory, which is obtained by neglecting the quantum vertex (\ref{eq:quantumvertex})~\cite{Aarts:2001yn,Berges:2007ym}. Therefore, we recover the fact that the leading infrared behavior of the scaling solutions is the same for the quantum and classical theory. Consequently, both the quantum and classical theory belong to a universality class characterized by the same critical exponents, which is well-established in thermal equilibrium. 

Using (\ref{eq:PiFclass}) and (\ref{eq:SigmaFclass}) the stationarity equation (\ref{eq:2PIstat})
can be written as
\begin{eqnarray}
&&\!\!\!\! \int_{qlr} \delta^{(d+1)}(p+l-q-r) \lambda_{{\rm eff},k}(p+l) \left\{\left[F_k(p) \rho_k(l) + \rho_k(p) F_k(l)\right] F_k(q) F_k(r) \right.
\nonumber\\
&&\!\!\!\! \left. - F_k(p) F_k(l) \left[F_k(q) \rho_k(r) + \rho_k(q) F_k(r)\right] \right\} \, = \, 0 \,,
\label{eq:fixedpointeq}
\end{eqnarray}
where we dropped an irrelevant overall factor $-\lambda/(18N)$.
Thermal equilibrium trivially solves (\ref{eq:fixedpointeq}) by virtue of the fluctuation dissipation relation (\ref{eq:flucdiss}). For scaling solutions this leads to the relation (\ref{eq:vacuumconst}) or (\ref{eq:constraint}) between the exponents $\kappa$, $z$ and $\eta$ as discussed in Sec.~\ref{sec:introfp}. It is less straightforward to determine the corresponding relations for nonthermal scaling solutions. Following Ref.~\cite{Berges:2008wm}, this may be efficiently achieved by integrating (\ref{eq:fixedpointeq}) over external spatial momentum ${\bf p}$ and  suitable coordinate transformations. In this way the problem is reduced to simple algebraic conditions for exponents.

The integrand of (\ref{eq:fixedpointeq}) contains a sum of four terms, which can be mapped onto each other up to proportionality factors using the scaling properties (\ref{eq:kscalingform}).
To map the second term in the integrand sum onto the first we employ for the frequencies the transformation 
\begin{equation}
l^0 \to \frac{p^0}{l^0} p^0 \, , \quad q^0 \to \frac{p^0}{l^0} q^0 \, , \quad r^0 \to \frac{p^0}{l^0} r^0 \, , \quad \left( p^0 \to \frac{p^0}{l^0} l^0 , \right)
\label{eq:freqtrans}
\end{equation}
which leads to the absolute value of the Jacobian determinant $\left( p^0/l^0 \right)^4$. For the spatial momenta we employ
\begin{equation}
{\bf l} \to \left(\frac{p^0}{l^0}\right)^{1/z} {\bf p} \, , \quad {\bf q} \to \left(\frac{p^0}{l^0}\right)^{1/z} {\bf q} \, , \quad {\bf r} \to \left(\frac{p^0}{l^0}\right)^{1/z} {\bf r} \, , \quad {\bf p} \to \left(\frac{p^0}{l^0}\right)^{1/z} {\bf l} \,.
\label{eq:spatialtrans}
\end{equation}  
Similarly, for the third term in the integrand sum we take $l^0 \to \left(p^0/r^0\right) q^0$,
$q^0 \to \left(p^0/r^0\right) p^0$, $r^0 \to \left(p^0/r^0\right) l^0$ and $p^0 \to \left(p^0/r^0\right) r^0$. Finally, the fourth term is transformed with $l^0 \to \left(p^0/q^0\right) r^0$,
$q^0 \to \left(p^0/q^0\right) l^0$, $r^0 \to \left(p^0/q^0\right) p^0$ and $p^0 \to \left(p^0/q^0\right) q^0$. Corresponding transformations are done for the spatial momenta analogous to (\ref{eq:spatialtrans}).

It remains to establish the scaling properties of $\lambda_{{\rm eff},k}(p)$ defined in (\ref{eq:effcoup}). The propagators (\ref{eq:kscalingform}) determine the behavior of the retarded/advanced self-energies defined in (\ref{eq:quantumPi}):
\begin{equation}
\Pi^{\mathrm R,A}_k(p^0,{\bf p}) = k^{-\Delta}\, \pi^{\mathrm R,A}\!\left( \frac{p^0}{k^z},\frac{|{\bf p}|}{k}\right) \, ,
\label{eq:scalingpira}
\end{equation}
where
\begin{equation}
\Delta = 4-d+\kappa-z-\eta \, ,
\label{eq:Delta}
\end{equation}
with scaling functions $\pi^{\mathrm R,A}$. The infrared behavior of (\ref{eq:effcoup}) 
is then for
\begin{equation}
\Delta > 0\, : \quad  \lambda_{{\rm eff},k}(p^0,{\bf p}) = k^{2 \Delta}\, l\! \left( \frac{p^0}{k^z},\frac{|{\bf p}|}{k}\right) \, ,
\label{eq:lscale}
\end{equation}
with scaling function $l$. For $\Delta \le 0$ the effective coupling becomes trivial
with $\lambda_{{\rm eff},k}(p) \simeq \lambda$, on which we comment below. 

Employing the above transformations to (\ref{eq:fixedpointeq}), and the scaling behavior for propagators (\ref{eq:kscalingform}) and effective vertex (\ref{eq:lscale}), one obtains for physical correlation functions at $k=0$:
\begin{eqnarray}
0 & = &\int_{{\bf p}qlr} \delta^{(d+1)}(p+l-q-r) 
\left[1 + \left(\frac{p^0}{l^0}\right)^{\tilde{\Delta}} - \left(\frac{p^0}{r^0}\right)^{\tilde{\Delta}} - \left(\frac{p^0}{q^0}\right)^{\tilde{\Delta}} \right]
\nonumber\\
&& \qquad \times \,  \lambda_{{\rm eff},k=0}(p+l) F_{k=0}(p) \rho_{k=0}(l) F_{k=0}(q) F_{k=0}(r)
\, , \label{eq:zeros}
\end{eqnarray}
where
\begin{equation}
\tilde{\Delta} \, = \, 1 + \frac{d -\kappa -\eta}{z}\, .
\end{equation}
Nonthermal solutions of (\ref{eq:zeros}) are manifest (I) for $\tilde{\Delta}=0$, which corresponds to (\ref{eq:nonthermalconstraint}) given in Sec.~\ref{sec:introfp}, and (II) for $\tilde{\Delta}=-1$. The second solution reads
\begin{equation}
{\rm (II):} \quad \kappa \,=\, - \eta + 2 z + d\, .
\label{eq:secondnonthermal}
\end{equation}
 
We finally comment on the case $\Delta \le 0$ in (\ref{eq:Delta}) for which the effective coupling becomes trivial. One can repeat the above steps with $\lambda_{{\rm eff},k}(p) \simeq \lambda$ and finds potential nonthermal solutions, which we will denote by an index (UV), given by
\begin{eqnarray}
{\rm (I):}  \quad \kappa^{\rm (UV)} & \! = \! & d - \frac{8}{3} + z^{\rm (UV)} + \frac{\eta^{\rm (UV)}}{3} \,,
\nonumber\\ 
{\rm (II):} \quad \kappa^{\rm (UV)} & \! = \! & d - \frac{8}{3} + \frac{4}{3} z^{\rm (UV)} + \frac{\eta^{\rm (UV)}}{3} \,.
\label{eq:UVsolutions}
\end{eqnarray}
However, plugging these solutions into (\ref{eq:Delta}) in order to verify the premise (\ref{eq:Delta}) yields (I) $\left(4-2\eta^{\rm (UV)}\right)/3 \le 0$ and (II) $\left(4-2\eta^{\rm (UV)}+z^{\rm (UV)}\right)/3 \le 0$. For typical values of the anomalous dimension and the dynamical scaling exponent this is excluded.

The type of scaling solutions (\ref{eq:UVsolutions}) can nevertheless play an important role for the physics of (wave) turbulence in classical-statistical systems~\cite{Zakharov}. To understand this, we note that $\lambda_{{\rm eff},k}(p) \simeq \lambda$ is physically realized for sufficiently large momentum $p$ according to (\ref{eq:effcoup}). The calculation becomes then perturbative and (\ref{eq:UVsolutions}) corresponds to the weak-coupling result for large $N$ if quantum corrections are neglected (i.e.\ $\Gamma^{{\tilde \phi}{\tilde \phi}{\tilde \phi}\phi}_{\Lambda}\equiv 0$ in (\ref{eq:quantumvertex})). The perturbative values $z^{\rm (UV)} = 1$ and $\eta^{\rm (UV)} = 0$ in (\ref{eq:UVsolutions}) lead, e.g., in $d=3$ to the well-known values $\kappa^{\rm (UV)} = 4/3$ and $5/3$~\cite{Zakharov}. 

It should be stressed that the UV fixed point solutions (\ref{eq:UVsolutions}) are derived neglecting quantum corrections. However, these become important in the high momentum regime, where occupation numbers are small. From the quantum self-energies (\ref{eq:NLOselfenergies}) we observe that the presence of quantum fluctuations prevents the UV scaling solutions (\ref{eq:UVsolutions}). In contrast, as discussed above quantum corrections do not alter the infrared fixed points (\ref{eq:nonthermalconstraint}) and (\ref{eq:secondnonthermal}) associated to critical phenomena far from equilibrium.

\section{Conclusions}
\label{sec:conclusions}

Our study reveals a hierarchy of fixed point solutions, which may be classified by the exponent $\kappa$ according to (\ref{eq:vacuumconst}), (\ref{eq:constraint}) and (\ref{eq:nonthermalconstraint}), or (\ref{eq:secondnonthermal}). The increasing complexity of $\kappa$ from vacuum and thermal equilibrium to nonequilibrium may be illustrated in terms of properties of occupation numbers. Vacuum carries vanishing occupation number and $\kappa$ is trivially determined by the anomalous dimension $\eta$. Thermal equilibrium is characterized by a Bose-Einstein distributed occupation number. This distribution is a function of frequency and not spatial momenta. As a consequence, in thermal equilibrium $\kappa$ depends in addition to $\eta$ also on the dynamical scaling exponent $z$, which characterizes scaling with respect to frequency. Nonthermal fixed points constitute the least constrained type of scaling solutions in this hierarchy. Nonequilibrium occupation numbers depend in general on frequencies as well as spatial momenta. We find that $\kappa$ indeed exhibits an explicit dependence on the number of spatial dimensions $d$, in addition to its possible dependencies via $\eta$ and $z$.         

These results are obtained using the functional renormalization group on a closed real-time path. This approach does not build in a fluctuation-dissipation relation, in contrast to Euclidean or complex-time formulations at nonzero temperature. As a consequence, it provides a common framework to address thermal as well as nonthermal fixed points. Though the derivation is in principle a straightforward generalization of standard Euclidean formulations, there are important differences which required a thorough discussion of the functional aspects of this approach. This includes basic relations such as (\ref{eq:FG}), which are crucial for practical calculations of exact equations for proper vertices by functional differentiation.  

Nonthermal fixed points represent basic properties of quantum field theories, just as thermal or vacuum fixed points, which provide an important means of classification. Further studies are required to reach a similar level of understanding as in equilibrium. This concerns, in particular, stability properties, which go beyond this work. It is remarkable that nonthermal infrared fixed points seem to be approached from a substantial class of nonequilibrium instability initial conditions without fine-tuning of parameters as reported in Ref.~\cite{Berges:2008wm}. These results indicate that due to critical slowing down the system stays for a long time in the vicinity of a nonthermal fixed point, acquiring its properties. Depending on how closely the fixed point is approached, subsequent deviations after sufficiently long times finally drive the system towards thermal equilibrium. In principle, one may directly tune to a nonthermal fixed point without ever deviating from it, which may be of measure zero. 

It would also be very interesting to further probe the predicted dependence of nonthermal fixed points on the dynamical exponent $z$ and on the dimensionality of space $d$, using lattice simulations in classical-statistical field theories in the same universality class. This could be used to verify (\ref{eq:scaling}) or whether more general scaling assumptions have to be made in order to explain the observed nonequilibrium properties. Nonrelativistic systems tend to have larger $z$ than
relativistic theories, which allows a simple variation of that parameter as well. Accordingly, a very fruitful direction could be the experimental realization of nonthermal fixed point dynamics in cold atomic gases, following a parametric resonance or spinodal nonequilibrium instability.\\

We thank Thomas Gasenzer, Alexander Rothkopf, and Jonas Schmidt for collaboration on related work, and Jan Pawlowski for very helpful discussions. This work is supported in part by the BMBF grant 06DA267, and by the DFG under contract SFB634. Part of this work was done during participation in the programme on Nonequilibrium Dynamics in Particle Physics and Cosmology at the Kavli Institute for Theoretical
Physics in Santa Barbara, supported by the National Science Foundation
under grant PHY05-51164.

\appendix

\section{Functional form of propagators}
\label{sec:funcform}

The identities (\ref{eq:FG}) relate second functional derivatives of $W_k[J,\tilde{J}] = W[J,\tilde{J},R_k^{\mathrm{R},\mathrm{A},F,{\tilde F}}]$ to those of $\Gamma_k[\phi,{\tilde \phi}] = \Gamma[\phi,{\tilde \phi};R_k^{\mathrm{R},\mathrm{A},F,{\tilde F}}]$. In the presence of the $R_k^{F,\mathrm{R},\mathrm{A},{\tilde F}}$ source terms the relations between the fields and  
$J=J_k[\phi,{\tilde \phi}]$ and ${\tilde J}={\tilde J}_k[\phi,{\tilde \phi}]$ are scale dependent. With (\ref{eq:Defphiphitilde}) and (\ref{eq:fields}) follows
\begin{eqnarray}
&&\!\!\!\! \delta^{(d+1)}(x-y) \, = \,  \frac{\delta J_k(x)}{\delta J_k(y)}
\, = \, \int_z \left(
\frac{\delta J_k(x)}{\delta \phi(z)} \frac{\delta \phi(z)}{\delta J_k(y)}
+ \frac{\delta J_k(x)}{\delta {\tilde \phi}(z)} \frac{\delta {\tilde \phi}(z)}{\delta J_k(y)} 
\right) 
\nonumber\\
&&\!\!\!\! \, = \, \int_z \left\{\left( - \frac{\delta^2 \Gamma_k}{\delta {\tilde \phi}(x) \delta \phi(z)}
- \frac{1}{2} R_k^{\mathrm R}(x,z) - \frac{1}{2} R_k^{\mathrm A}(z,x) \right) 
\frac{\delta^2 W_k}{\delta {\tilde J}(z) \delta J(y)} \right.
\nonumber\\
&&\!\!\!\! \left.  \qquad + \, \left( - \frac{\delta^2 \Gamma_k}{\delta {\tilde \phi}(x) \delta {\tilde \phi}(z)}
- R_k^F(x,z) \right) 
\frac{\delta^2 W_k}{\delta J(z) \delta J(y)} \right\}\, ,
\end{eqnarray}
which reads in compact matrix notation using (\ref{eq:RARR}):
\begin{equation}
\left(\Gamma_k^{{\tilde \phi}\phi} + R_k^{\mathrm R} \right) W_k^{{\tilde J}J} +
\left(\Gamma_k^{{\tilde \phi}{\tilde \phi}} + R_k^F \right) W_k^{JJ} \, = \, - {\mathbf 1} \, .
\label{eq:GW1}
\end{equation}
Similarly, the other combinations of functional derivatives with $J$ and ${\tilde J}$ can be written as
\begin{eqnarray}
\left(\Gamma_k^{\phi\phi} + R_k^{\tilde F} \right) W_k^{{\tilde J}{\tilde J}} +
\left(\Gamma_k^{\phi{\tilde \phi}} + R_k^{\mathrm A} \right) W_k^{J{\tilde J}} 
& \! = \! & - {\mathbf 1} \, ,
\\
\left(\Gamma_k^{{\tilde \phi}\phi} + R_k^{\mathrm R} \right) W_k^{{\tilde J}{\tilde J}} +
\left(\Gamma_k^{{\tilde \phi}{\tilde \phi}} + R_k^{F} \right) W_k^{J{\tilde J}} 
& \! = \! & 0 \, ,
\\
\left(\Gamma_k^{\phi\phi} + R_k^{\tilde F} \right) W_k^{{\tilde J}J} +
\left(\Gamma_k^{\phi{\tilde \phi}} + R_k^{\mathrm A} \right) W_k^{JJ} 
& \! = \! & 0 \, .
\label{eq:GW4}
\end{eqnarray}
The above equations can be solved for $W_k^{{\tilde J}{\tilde J}}$, $W_k^{J{\tilde J}}$, $W_k^{{\tilde J}J}$ and $W_k^{JJ}$ in terms of the second derivatives of $\Gamma_k$ and $R_k^{\mathrm{R},\mathrm{A},F,{\tilde F}}$, which yields (\ref{eq:FG}) using the definitions (\ref{eq:propdef}).

\section{Flow equations for a sharp cutoff}
\label{sec:sharpcutoff}

The flow equations in Sec.~\ref{sec:dia} may be considerably simplified using a sharp cutoff function $R^{\mathrm R}_k(p)$, which diverges for $p^2 < k^2$ and vanishes for momenta larger than $k^2$. In Fourier space, the propagators (\ref{eq:dia}) can then be written as
\begin{equation}
G^{\mathrm R}_k(p) = - \left(\Gamma^{{\tilde \phi} \phi}_k\right)^{-1} \, \Theta (p^2 - k^2) \quad , \quad
G^{\mathrm A}_k(p) = - \left(\Gamma^{\phi {\tilde \phi}}_k\right)^{-1} \, \Theta (p^2 - k^2). 
\label{eq:sharpprop}
\end{equation}
Flow equations for two-point functions given by (\ref{eq:exactG2dia}), or (\ref{eq:G2Fourier}), contain  in matrix notation
\begin{eqnarray}
\lefteqn{G^{\mathrm R}_k \dot{R}^{\mathrm R}_k G^{\mathrm R}_k
= \left(\Gamma^{{\tilde \phi} \phi}_k + R^{\mathrm R}_k\right)^{-1}
\dot{R}^{\mathrm R}_k \left(\Gamma^{{\tilde \phi} \phi}_k + R^{\mathrm R}_k\right)^{-1}} 
\nonumber\\
&\!=\!& - 2 k^2 \frac{\partial}{\partial k^2} \left(\Gamma^{{\tilde \phi} \phi}_k + R^{\mathrm R}_k\right)^{-1} - \left(\Gamma^{{\tilde \phi} \phi}_k + R^{\mathrm R}_k\right)^{-1}
\dot{\Gamma}^{{\tilde \phi} \phi}_k \left(\Gamma^{{\tilde \phi} \phi}_k + R^{\mathrm R}_k\right)^{-1} .
\label{eq:sharp1}
\end{eqnarray} 
Using (\ref{eq:sharpprop}) the last expression of (\ref{eq:sharp1}) can be written as
\begin{eqnarray}
\lefteqn{- 2 k^2 \frac{\partial}{\partial k^2} \left\{ \Theta(p^2 - k^2) \left(\Gamma^{{\tilde \phi} \phi}_k\right)^{-1} \right\} - \Theta(p^2 - k^2) \left(\Gamma^{{\tilde \phi} \phi}_k\right)^{-1} \dot{\Gamma}^{{\tilde \phi} \phi}_k \left(\Gamma^{{\tilde \phi} \phi}_k\right)^{-1}}  
\nonumber\\
&=& - 2 k^2 \frac{\partial}{\partial k^2} \left\{ \Theta(p^2 - k^2) \left(\Gamma^{{\tilde \phi} \phi}_k\right)^{-1}\right\} + \Theta(p^2 - k^2) 2 k^2 \frac{\partial}{\partial k^2} \left(\Gamma^{{\tilde \phi} \phi}_k\right)^{-1}
\nonumber\\
&=& 2 k^2 \delta(p^2 - k^2) \left(\Gamma^{{\tilde \phi} \phi}_k\right)^{-1} . 
\label{eq:reduce}
\end{eqnarray}
Similarly,
\begin{equation}
G^{\mathrm A}_k \dot{R}^{\mathrm A}_k G^{\mathrm A}_k
= 2 k^2 \delta(p^2 - k^2) \left(\Gamma^{\phi{\tilde \phi}}_k\right)^{-1} .
\label{eq:sharp2}
\end{equation}
In the flow equations for two-point functions also appears
$G^{\mathrm R}_k \dot{R}^{\mathrm R}_k i F_k = G^{\mathrm R}_k \dot{R}^{\mathrm R}_k G^{\mathrm R}_k \Gamma^{{\tilde \phi} {\tilde \phi}}_k G^{\mathrm A}_k$ and one may be tempted to evaluate this accordingly using (\ref{eq:reduce}). However, with (\ref{eq:sharpprop}) this naive replacement would lead to ill-defined expressions $\sim \delta(p^2-k^2) \Theta(p^2-k^2)$. Instead, a proper treatment takes into account that only the following well-defined sum appears:
\begin{eqnarray}
\lefteqn{G^{\mathrm R}_k \dot{R}^{\mathrm R}_k i F_k + i F_k \dot{R}^{\mathrm A}_k G^{\mathrm A}_k}
\nonumber\\
&\!=\!& 2 k^2 \frac{\partial}{\partial k^2} \left\{ \left(\Gamma^{{\tilde \phi} \phi}_k + R^{\mathrm R}_k\right)^{-1} \Gamma^{{\tilde \phi} {\tilde \phi}}_k \left(\Gamma^{\phi {\tilde \phi}}_k + R^{\mathrm A}_k\right)^{-1} \right\}
\nonumber\\
&\!+\!& \left(\Gamma^{{\tilde \phi} \phi}_k + R^{\mathrm R}_k\right)^{-1}
\dot{\Gamma}^{{\tilde \phi} \phi}_k \left(\Gamma^{{\tilde \phi} \phi}_k + R^{\mathrm R}_k\right)^{-1}\Gamma^{{\tilde \phi} {\tilde \phi}}_k \left(\Gamma^{\phi {\tilde \phi}}_k + R^{\mathrm A}_k\right)^{-1}
\nonumber\\
&\!+\!& \left(\Gamma^{{\tilde \phi} \phi}_k + R^{\mathrm R}_k\right)^{-1}\dot{\Gamma}^{{\tilde \phi} {\tilde \phi}}_k \left(\Gamma^{\phi {\tilde \phi}}_k + R^{\mathrm A}_k\right)^{-1}
\nonumber\\
&\!+\!& \left(\Gamma^{{\tilde \phi} \phi}_k + R^{\mathrm R}_k\right)^{-1}
\Gamma^{{\tilde \phi} {\tilde \phi}}_k \left(\Gamma^{\phi {\tilde \phi}}_k + R^{\mathrm A}_k\right)^{-1}\dot{\Gamma}^{\phi {\tilde \phi}}_k \left(\Gamma^{\phi {\tilde \phi}}_k + R^{\mathrm A}_k\right)^{-1}
\nonumber\\
&\!=\!& - 2 k^2 \delta(p^2 - k^2) \left(\Gamma^{{\tilde \phi} \phi}_k\right)^{-1} \Gamma^{{\tilde \phi} {\tilde \phi}}_k \left(\Gamma^{\phi {\tilde \phi}}_k \right)^{-1}. 
\label{eq:sharp3}
\end{eqnarray} 
In order to integrate flow equations for four-point functions, we employ
\begin{eqnarray}
&&\!\!\!\!\left(G^{\mathrm R}_k \dot{R}^{\mathrm R}_k G^{\mathrm R}_k\right)(p-q) \, \Gamma^{(4)}_k G^{\mathrm R}_k(q) \, \Gamma^{(4)}_k +G^{\mathrm R}_k(p-q) \, \Gamma^{(4)}_k \left(G^{\mathrm R}_k \dot{R}^{\mathrm R}_k G^{\mathrm R}_k\right)(q) \, \Gamma^{(4)}_k\nonumber\\
&&\!\!\!\! =\, - 2 k^2 \left\{\delta\left((p-q)^2 -k^2\right) \, \Theta(q^2-k^2) +\delta(q^2 -k^2) \, \Theta\left((p-q)^2-k^2\right)\right\} \nonumber\\
&&\!\!\!\!\times\, \left(\Gamma^{{\tilde \phi} \phi}_k\right)^{-1}\!\!(p-q) \,\, \Gamma^{(4)}_k \left(\Gamma^{{\tilde \phi} \phi}_k\right)^{-1}\!\!(q) \,\, \Gamma^{(4)}_k\nonumber\\
&&\!\!\!\!=\, - 2 k^2 \left\{\delta\left((p-q)^2 -k^2\right) \, \Theta(-p^2+2pq) +\delta(q^2 -k^2) \, \Theta(p^2-2pq)\right\} \nonumber\\
&&\!\!\!\!\times\, \left(\Gamma^{{\tilde \phi} \phi}_k\right)^{-1}\!\!(p-q) \,\, \Gamma^{(4)}_k \left(\Gamma^{{\tilde \phi} \phi}_k\right)^{-1}\!\!(q) \,\, \Gamma^{(4)}_k .
\label{eq:sharp4}
\end{eqnarray}
Similarly, 
\begin{eqnarray}
&&\!\!\!\!\left(G^{\mathrm R}_k \dot{R}^{\mathrm R}_k G^{\mathrm R}_k\right)(p-q) \, \Gamma^{(4)}_k G^{\mathrm A}_k(q) \, \Gamma^{(4)}_k +G^{\mathrm R}_k(p-q) \, \Gamma^{(4)}_k \left(G^{\mathrm A}_k \dot{R}^{\mathrm A}_k G^{\mathrm A}_k\right)(q) \, \Gamma^{(4)}_k \nonumber\\
&&\!\!\!\! =\, - 2 k^2 \left\{\delta\left((p-q)^2 -k^2\right) \, \Theta(q^2-k^2) +\delta(q^2 -k^2) \, \Theta\left((p-q)^2-k^2\right)\right\} \nonumber\\
&&\!\!\!\!\times\, \left(\Gamma^{{\tilde \phi} \phi}_k\right)^{-1}\!\!(p-q) \,\, \Gamma^{(4)}_k \left(\Gamma^{\phi{\tilde \phi}}_k\right)^{-1}\!\!(q) \,\, \Gamma^{(4)}_k.
\label{eq:sharp5}
\end{eqnarray}
Expressions for four-point flow equations involving $iF$ read
\begin{eqnarray}
&&\!\!\!\!\left(G^{\mathrm R}_k \dot{R}^{\mathrm R}_k G^{\mathrm R}_k\right)(p-q) \, \Gamma^{(4)}_k iF_k(q) \, \Gamma^{(4)}_k
+ G^{\mathrm R}_k(p-q) \, \Gamma^{(4)}_k \left(G^{\mathrm R}_k \dot{R}^{\mathrm R}_k iF_k \right)(q) \, \Gamma^{(4)}_k \nonumber\\
&&\!\!\!\! +\, G^{\mathrm R}_k(p-q) \, \Gamma^{(4)}_k \left(iF_k \dot{R}^{\mathrm A}_k G^{\mathrm A}_k\right)(q) \, \Gamma^{(4)}_k \nonumber\\
&&\!\!\!\! =\, 2 k^2 \left\{\delta\left((p-q)^2 -k^2\right) \, \Theta(q^2-k^2) +\delta(q^2 -k^2) \, \Theta\left((p-q)^2-k^2\right)\right\} \nonumber\\
&&\!\!\!\!\times\, \left(\Gamma^{{\tilde \phi} \phi}_k\right)^{-1}\!\!(p-q) \,\, \Gamma^{(4)}_k \left(\left(\Gamma^{{\tilde \phi} \phi}_k\right)^{-1} \Gamma^{{\tilde \phi} {\tilde \phi}}_k \left(\Gamma^{\phi {\tilde \phi}}_k\right)^{-1}\right)(q) \,\, \Gamma^{(4)}_k,
\label{eq:sharp6}
\end{eqnarray}
and
\begin{eqnarray}
&&\!\!\!\!\left(G^{\mathrm A}_k \dot{R}^{\mathrm A}_k G^{\mathrm A}_k\right)(p-q) \, \Gamma^{(4)}_k \, iF_k(q) \, \Gamma^{(4)}_k
+ G^{\mathrm A}_k (p-q) \, \Gamma^{(4)}_k \left(G^{\mathrm R}_k \dot{R}^{\mathrm R}_k iF_k\right)(q) \, \Gamma^{(4)}_k \nonumber\\
&&\!\!\!\!+\, G^{\mathrm A}_k (p-q) \, \Gamma^{(4)}_k \left(iF_k \dot{R}^{\mathrm A}_k G^{\mathrm A}_k\right)(q) \, \Gamma^{(4)}_k  \nonumber\\
&&\!\!\!\! =\, 2 k^2 \left\{\delta\left((p-q)^2 -k^2\right) \, \Theta(q^2-k^2) +\delta(q^2 -k^2) \, \Theta\left((p-q)^2-k^2\right)\right\} \nonumber\\
&&\!\!\!\! \times\, \left(\Gamma^{\phi{\tilde \phi}}_k\right)^{-1}\!\!(p-q) \,\, \Gamma^{(4)}_k \left(\left(\Gamma^{{\tilde \phi} \phi}_k\right)^{-1} \Gamma^{{\tilde \phi} {\tilde \phi}}_k \left(\Gamma^{\phi {\tilde \phi}}_k\right)^{-1}\right)(q) \,\, \Gamma^{(4)}_k,
\label{eq:sharp7}
\end{eqnarray}
as well as
\begin{eqnarray}
&&\!\!\!\!\left( G^{\mathrm R}_k \dot{R}^{\mathrm R}_k iF_k \,+\,iF_k \dot{R}^{\mathrm A}_k G^{\mathrm A}_k \right)(p-q) \, \Gamma^{(4)}_k iF_k (q) \, \Gamma^{(4)}_k \nonumber\\
&&\!\!\!\! +\, iF_k(p-q) \, \Gamma^{(4)}_k \left(G^{\mathrm R}_k \dot{R}^{\mathrm R}_k iF_k \,+\,iF_k \dot{R}^{\mathrm A}_k G^{\mathrm A}_k \right)(q) \, \Gamma^{(4)}_k \nonumber\\
&&\!\!\!\! =\, - 2 k^2 \left\{\delta\left((p-q)^2 -k^2\right) \, \Theta(q^2-k^2) +\delta(q^2 -k^2) \, \Theta\left((p-q)^2-k^2\right)\right\} \nonumber\\
&&\!\!\!\! \times\, \left(\left(\Gamma^{{\tilde \phi} \phi}_k\right)^{-1} \Gamma^{{\tilde \phi} {\tilde \phi}}_k \left(\Gamma^{\phi {\tilde \phi}}_k\right)^{-1}\right)(p-q) \,\, \Gamma^{(4)}_k \left(\left(\Gamma^{{\tilde \phi} \phi}_k\right)^{-1} \Gamma^{{\tilde \phi} {\tilde \phi}}_k \left(\Gamma^{\phi {\tilde \phi}}_k\right)^{-1}\right)(q) \,\, \Gamma^{(4)}_k .\quad
\nonumber\\
\label{eq:sharp8}
\end{eqnarray}
As exemplified in the last equation of (\ref{eq:sharp4}), the two theta functions contributing to the flow equations for $\Gamma_k^{(4)}$ are both independent of $k$ and have the same argument but with opposite sign such that they sum up to unity after $k$-integration.

\end{document}